\documentclass[12pt,a4paper]{article}
\usepackage{jheppub}  
\usepackage{amssymb} 
\usepackage{amsmath}
\usepackage{mathtools}
\usepackage{amsfonts}    
\usepackage{dsfont}
\usepackage{pdfpages}
\usepackage{verbatim}
\usepackage{tensor}
\usepackage{mathrsfs}
\usepackage[mathscr]{euscript}
\usepackage{tikz}

\newcommand{\ba}{\begin{align}}

\newcommand{\be}{\begin{equation}}
\newcommand{\ee}{\end{equation}}
\def\bd{\begin{tikzpicture}}
\def\ed{\end{tikzpicture}}

\allowdisplaybreaks[1]

\title{String theory on $\boldsymbol{\text{AdS}_{\mathbf{3}}}$ and the symmetric orbifold of Liouville theory}

\author{Lorenz Eberhardt and Matthias R.\ Gaberdiel} 
\affiliation{Institut f\"ur Theoretische Physik, ETH Zurich, \\
CH-8093 Z\"urich, Switzerland}
\emailAdd{eberhardtl@itp.phys.ethz.ch, gaberdiel@itp.phys.ethz.ch}

\abstract{For string theory on AdS$_3$ with pure NS-NS flux a complete set of DDF operators is constructed, from which one can read off the symmetry algebra of the spacetime CFT. Together with an analysis of the spacetime spectrum, this allows us to show that the CFT dual of superstring theory on ${\rm AdS}_3 \times {\rm S}^3 \times \mathbb{T}^4$ for generic NS-NS flux is the symmetric orbifold of $({\cal N}=4\ \text{Liouville theory})\times \mathbb{T}^4$. For the case of minimal flux ($k=1$), the Liouville factor disappears, and we just obtain the symmetric orbifold of $\mathbb{T}^4$, thereby giving further support to a previous claim. We also show that a similar analysis can be done for bosonic string theory on AdS$_3 \times X$.}

\begin{document}

\maketitle

\makeatletter
\g@addto@macro\bfseries{\boldmath}
\makeatother

\section{Introduction} \label{sec:intro}
Holography on $\mathrm{AdS}_3$ backgrounds has provided us with a useful tool to test ideas about the AdS/CFT correspondence in a much more controlled setting than in the higher dimensional cases \cite{Maldacena:1997re}. On the one hand, string theory on $\mathrm{AdS}_3$ backgrounds is dual to two-dimensional CFTs, which are often exactly solvable. On the other hand, string theory on $\mathrm{AdS}_3$ backgrounds admits an exact solution in terms of a worldsheet description employing WZW models \cite{Maldacena:2000hw, Maldacena:2000kv, Maldacena:2001km}. This worldsheet description corresponds to a pure NS-NS flux background.

It has often been asserted in the literature that the pure NS-NS background is `singular', in the sense that it features a continuum in the spectrum \cite{Seiberg:1999xz, Larsen:1999uk, Eberhardt:2018vho}. 
This continuum is associated to strings which can reach the boundary of $\mathrm{AdS}_3$ at a finite cost of energy -- the so-called long strings.
Hence, the dual CFT necessarily also possesses such a continuum of states, which renders the vacuum of the dual CFT non-normalisable.

It was shown in \cite{Eberhardt:2018ouy} that at least in the supersymmetric setting on $\mathrm{AdS}_3 \times \mathrm{S}^3 \times \mathbb{T}^4$, something special happens if the NS-NS flux takes its smalles value ($k=1$), in which case the strings become tensionless. In this case, the continuum vanishes completely from the spectrum due to various shortening conditions on the worldsheet. As a result the $k=1$ background is dual to a bona fide CFT without a continuum. By matching the complete partition function and the fusions rules, strong evidence was given in \cite{Eberhardt:2018ouy} that this CFT is in fact the much-discussed symmetric orbifold of $\mathbb{T}^4$, see e.g.\ \cite{David:2002wn}. This yields an example of an AdS/CFT duality, in which both sides of the duality are exactly solvable.
\medskip

The purpose of this paper is twofold. First, we provide further evidence for the picture advocated in  \cite{Eberhardt:2018ouy} by showing that not only the spectrum matches, but that we can also reproduce the algebraic structure of the dual CFT, in particular, the commutation relations of the spectrum generating fields. This goes a long way towards proving the duality in this case. The other main result is to show that a large part of the analysis of \cite{Eberhardt:2018ouy} can also be done for $k>1$, and that this allows us to make a convincing conjecture for the CFT dual in the more general case: for superstring theory on $\mathrm{AdS}_3 \times \mathrm{S}^3 \times \mathbb{T}^4$ with $k$ units of NS-NS flux, we propose that the dual CFT  is the symmetric orbifold, 
\be 
\text{Sym}^N\left(\Bigl[\mathcal{N}=4\text{ Liouville with }c=6(k-1)\Bigr] \oplus\ \mathbb{T}^4\right)\ ,\label{introduction symmetric product susy}
\ee
in the large $N$ limit, 
see also \cite{Yu:1998qw, Hosomichi:1998be, Hikida:2000ry, Argurio:2000tb, Giveon:2005mi} for related work.
\medskip

The main idea behind elucidating the algebraic structure of the dual CFT is to construct a complete set of `DDF operators' \cite{DelGiudice:1971yjh} on the world-sheet. These operators commute with the physical state conditions and hence act on the space of physical states. In the context of AdS$_3$, the construction of these DDF operators was already pioneered some time ago \cite{Giveon:1998ns}, see also \cite{Kutasov:1999xu, deBoer:1998gyt, Giveon:2001up} for subsequent developments. We complete and extend this analysis, and then use it to show that the symmetry algebra of the spacetime CFT is indeed the chiral algebra of the above symmetric orbifold.


The worldsheet WZW model describing $\mathrm{AdS}_3$ is based on the affine algebra $\mathfrak{sl}(2,\mathds{R})_k$, and the full spectrum of the theory consists of a certain family of discrete and continuous representations of $\mathfrak{sl}(2,\mathds{R})$, together with their spectrally flowed images. The states from the $w$ spectrally flowed sector correspond to strings which wind asymptotically $w$ times around the boundary of $\mathrm{AdS}_3$. The key observation of our analysis is to note that the moding of the DDF operators depends critically on the spectral flow sector they act on. For example, while the central charge of the spacetime Virasoro algebra \`a la Brown-Henneaux \cite{Brown:1986nw} apparently equals $c=6kw$ in the $w$'th flowed sector \cite{Giveon:1998ns}, the modes of the spacetime Virasoro algebra are actually allowed to take values in $\frac{1}{w}\, \mathds{Z}$ in this sector.\footnote{This is the case for the continuous representations on the world-sheet; for the discrete representations the situation is more complex, see Section~\ref{subsec:discretereps} and below.} This is very reminiscent of the $w$-cycle twisted sector of a symmetric orbifold, and we show that this interpretation is indeed correct.  Furthermore, we show that a similar argument applies to all the other DDF operators (not just the spacetime Virasoro generators). 

\medskip

We will first exemplify the construction for bosonic string theory on $\mathrm{AdS}_3$, where many technical complications are absent. We will show that the spectrum generating algebra of the spacetime CFT for the background $\mathrm{AdS}_3 \times X$ is that of the symmetric orbifold of the Virasoro algebra times the chiral algebra of the (arbitrary) internal CFT $X$. We also determine the representations of this algebra that actually appear in the spacetime spectrum, and this leads to the conclusion that the spacetime CFT is the large $N$ limit of
\be
\text{Sym}^N\left(\left[\text{Liouville with }c=1+\frac{6(k-3)^2}{k-2}\right] \times X\right)\ , \label{introduction symmetric product}
\ee
i.e.\ the symmetric orbifold of Liouville theory times the internal CFT. Here, the continuum of Liouville theory arises precisely from the continuum of long string excitations in the bulk.

While this proposal nicely encompasses all the long strings in $\mathrm{AdS}_3$, it does not account for the short string solutions (that arise from the discrete representations on the world-sheet). These describe non-normalizable states of the dual CFT and therefore do not explicitly appear in the CFT spectrum (in the same way as the vacuum does not appear in Liouville theory), see also \cite{Maldacena:2001km} for a related discussion. 

We also analyse in detail the supersymmetric background $\mathrm{AdS}_3 \times \mathrm{S}^3 \times \mathbb{T}^4$, in which case the analogous conclusion to \eqref{introduction symmetric product} is (\ref{introduction symmetric product susy}). We pay particular attention to the case of $k=1$, for which the Liouville part becomes trivial, thus explaining the absence of the long string continuum from this viewpoint. Furthermore for $k=1$, there are no discrete representations on the worldsheet \cite{Eberhardt:2018ouy}, and hence the analysis is complete. 
\bigskip

The paper is organised as follows. In Section~\ref{sec:bosonic} we develop the theory in the technically simpler setting of bosonic string theory on $\mathrm{AdS}_3 \times X$. We explain how to define the DDF operators and discuss their interpretation in detail. We also comment on the role of the discrete representations (short strings) in Section~\ref{subsec:discretereps}.
We then move on to the supersymmetric setting in Section~\ref{sec:review of strings}. It will be convenient to work at least partially in the hybrid formalism \cite{Berkovits:1999im}, as it makes spacetime supersymmetry manifest and can be used to define the $k=1$ worldsheet theory. We explain carefully how the degrees of freedom of the NS-R formalism can be rewritten in terms of hybrid fields, and how this gives rise to the $\mathfrak{psu}(1,1|2)_k$ WZW model that is discussed in detail in Section~\ref{sec:psu112k WZW model}. With these preparations at hand, we define the DDF operators in Section~\ref{sec:spacetime symmetry algebra} and compute their algebra. Finally, we identify the dual CFT of eq.~\eqref{introduction symmetric product susy} in Section~\ref{sec:symmetric product orbifold}. We end with a discussion of our findings in Section~\ref{sec:discussion}.
Appendix~\ref{app:higher spins} contains some details about the definition of the DDF algebra for general world-sheet CFTs, whereas Appendices~\ref{app:OPEs} and \ref{app:freeconv} mainly fix our conventions.

\section{Bosonic strings on \texorpdfstring{$\boldsymbol{\mathrm{AdS}_3}$}{AdS3}}\label{sec:bosonic}

As a warm-up to the technically more complex situation with supersymmetry, let us begin by analysing 
bosonic string theory on 
\be 
\mathrm{AdS}_3 \times X\ .
\ee

\subsection{The \texorpdfstring{$\mathfrak{sl}(2,\mathds{R})_k$}{sl(2,R)k} WZW model and its free field realisation}

The $\mathrm{AdS}_3$ part of the background can be described by an $\mathfrak{sl}(2,\mathds{R})_k$ WZW-model, and criticality of the background then imposes
\be 
\frac{3k}{k-2}+c_X=26\ . \label{eq:criticality}
\ee
In order to describe the vertex operators of the background, it is useful to employ the 
Wakimoto representation of $\mathfrak{sl}(2,\mathds{R})_{k}$. Let us introduce a pair of bosonic ghosts with $\lambda=1$ (see Appendix~\ref{app:freeconv} for our conventions)
\be 
\beta(z)\gamma(w) \sim -\frac{1}{z-w}\ , \label{eq:betagamma OPE}
\ee
as well as a free boson
\be 
\partial \Phi(z)\partial\Phi(w) \sim-\frac{1}{(z-w)^2}\ . \label{eq:PhiPhi OPE}
\ee
The free boson has background charge $Q = \sqrt{\frac{1}{k-2}}$ so that the total central charge equals 
\be
c = 2 + 1 + \frac{6}{k-2} = \frac{3k}{k-2} = c\bigl(\mathfrak{sl}(2,\mathds{R})_{k} \bigr) \ . 
\ee
We then have the Wakimoto representation
\begin{subequations}
\begin{align}
J^+&=\beta\ , \label{Wakib a}\\
J^3&=\sqrt{\tfrac{k-2}{2}}\partial\Phi+(\beta\gamma)\ , \label{Wakib b}\\
J^-&=\sqrt{2(k-2)} (\partial\Phi \gamma)+(\beta\gamma\gamma)-k\partial\gamma\ . \label{Wakib c}
\end{align}
\end{subequations}
We should mention that treating $\Phi$ as a free field is only adequate near the boundary of $\mathrm{AdS}_3$, see \cite{Giveon:1998ns} for a discussion. More generally, it should be understood as a Liouville field. However, since we are mainly interested in constructing the boundary CFT, the above description will be sufficient for our purposes.

\subsection{Vertex operators} \label{subsec:vertex operatorsb}

The spectrum of the WZW model consists of affine highest weight representations of $\mathfrak{sl}(2,\mathds{R})_{k}$ (whose ground states we label by $|j,m\rangle$), together with their spectrally flowed images. In the above free field realisation of $\mathfrak{sl}(2,\mathds{R})_{k}$, the affine highest weight state  $|j,m\rangle$ is described by 
\be\label{jmfree}
|j,m\rangle = \oint \mathrm{d}z\ \gamma^{-j-m} \mathrm{e}^{\, j\sqrt{\frac{2}{k-2}}\Phi}(z)\, |0\rangle \ . 
\ee
Indeed, this state is annihilated by the positive modes of \eqref{Wakib a}--\eqref{Wakib c}, and it transforms under the zero mode algebra $\mathfrak{sl}(2,\mathds{R})$
in a representation with Casimir $\mathcal{C}=-j(j-1)$
\begin{subequations}\label{2.8}
\begin{align}
J^+_0|j,m \rangle&=(m+j)|j,m+1 \rangle\ , \\
J^3_0 |j,m \rangle&=m|j,m \rangle\ , \\
J^-_0 |j,m \rangle&=(m-j) |j,m-1 \rangle\ .
\end{align}
\end{subequations}
Here $j$ can either be real, in which case we are dealing with a discrete representation, and then either $m-j \in \mathds{Z}_{\ge 0}$ or $m+j \in \mathds{Z}_{\le 0}$ so that the representation truncates. Alternatively, $j$ can also take the value $j=\tfrac{1}{2}+i p$ where $p \in \mathds{R}$, in which case the Casimir is still real. In this case, $m \in \mathds{Z}+\lambda$ for an arbitrary $\lambda \in \mathds{R}/\mathds{Z}$ and the representation does not truncate (except for $p=\tfrac{1}{2}$ and $\lambda=\tfrac{1}{2}$). These are the continuous representations of $\mathfrak{sl}(2,\mathds{R})$, which we shall denote by $\mathscr{C}^j_\lambda$. In the following we shall be treating both cases simultaneously. 

We should mention that the continuous representations with $j=\tfrac{1}{2}+ip$ correspond to the usual representations of the Liouville field $\Phi$, whereas the discrete representations with $j$ real describe non-normalisable representations. Indeed, the conformal dimension of the highest weight state $|j,m\rangle$ is 
\be\label{worldsheetLiouville}
h\bigl( |j,m\rangle \bigr) = - \frac{j(j-1)}{k-2} = - \alpha (\alpha - Q) \ , \qquad Q = \frac{1}{\sqrt{k-2}} \ ,
\ee
where we have written $j=\frac{\alpha}{Q}$, so that $j=\tfrac{1}{2}+ip$ translates into $\alpha = \frac{Q}{2} + i \hat{p}$ with $\hat{p}=Qp$. This will be a common theme throughout this paper, namely that the continuous representations on the world-sheet behave much better than the discrete (non-normalisable) representations. Finally, we should mention  that the representations with $j$ and $1-j$ are in fact identified thanks to the reflection formula of Liouville theory \cite{Teschner:2001rv}.

\subsubsection{Spectral flow} \label{subsec:spectral flow}
We will also need the behaviour of the Wakimoto representation under the spectral flow automorphism $\sigma$ of $\mathfrak{sl}(2,\mathds{R})_k$, which transforms the fields according to
\begin{subequations}
\begin{align}
\sigma^w(J^\pm)(z)&= J^\pm(z) z^{\mp w}\ , \\
\sigma^w(J^3)(z)&= J^3(z)+\tfrac{kw}{2z}\ .
\end{align}
\end{subequations}
This forces the fields of the Wakimoto representation to transform as
\begin{subequations}
\begin{align}
\sigma^w(\beta)(z)&=\beta(z)z^{-w}\ , \\
\sigma^w(\gamma)(z)&=\gamma(z)z^w\ , \label{wgam bos} \\
\sigma^w(\partial\Phi)(z)&=\partial \Phi(z)+\sqrt{\tfrac{k-2}{2}} \tfrac{w}{z}\ .
\end{align}
\end{subequations}
Applying spectral flow to the affine highest weight state (\ref{jmfree}) thus leads to 
\be 
|j,m,w \rangle=\oint \mathrm{d}z\ z^{-mw}\gamma^{-j-m} \mathrm{e}^{\, j\sqrt{\frac{2}{k-2}}\Phi}(z)|0\rangle\ . \label{eq:sl2R highest weight states spectrally flowed}
\ee

\subsection{The DDF operators}\label{sec:DDF}

The next step of our analysis consists of constructing the DDF operators  \cite{DelGiudice:1971yjh}, i.e.\ the spectrum generating operators of the spacetime CFT from the world-sheet. 

\subsubsection{The Virasoro algebra}

The most generic spacetime generators are the Virasoro generators that can be constructed following 
\cite{Giveon:1998ns}. They can be formulated purely in terms of the $\mathfrak{sl}(2,\mathds{R})_k$ worldsheet currents; this reflects that the conformal symmetry of the spacetime CFT arises from the AdS$_3$ factor. To this end we make the ansatz 
\be 
\mathcal{L}_m=\oint \mathrm
{d}z  \left(\alpha_3(m) \gamma^m J^3+\alpha_+(m) \gamma^{m+1} J^++\alpha_-(m) \gamma^{m-1} J^-\right)(z)\ ,
\ee
where we have introduced the curly $\mathcal{L}$ symbol in order to distinguish the spacetime Virasoro algebra from the world-sheet Virasoro generators. Here the exponents of $\gamma$ are chosen such that the $\mathfrak{sl}(2,\mathds{R})$-charges are homogeneous. In order for $\mathcal{L}_m$ to be a DDF operator, it has to satisfy the following requirements:
\begin{enumerate}
\item[(i)] It has to commute with the physical state conditions, i.e.\ the string theory BRST operator. In the context of bosonic string theory this requirement is equivalent to the condition that the integrand is a primary field of conformal dimension 1. A direct computation shows that this requires
\be 
m \alpha_3(m)+(m+1)\alpha_+(m) +(m-1)\alpha_-(m)=0\ .
\ee
\item[(ii)] It must not be BRST trivial, i.e.\ the integrand must not be a Virasoro descendant. The BRST trivial combination of the integrand is 
\be \label{BRSTtriv}
\gamma^m J^3-\tfrac{1}{2} \gamma^{m+1} J^+-\tfrac{1}{2}\gamma^{m-1} J^-=\tfrac{k}{2}\,  \gamma^{m-1}\partial \gamma \ ,
\ee
which is a total derivative unless $m=0$ (in which case this will only shift the zero mode $\mathcal{L}_0$ by the spacetime identity \eqref{Idef}, see the discussion in Section~\ref{subsec:identity operator}). Thus we are free to redefine the DDF operator by adding a multiple of (\ref{BRSTtriv}).
\item[(iii)] Finally, we may require that the three global (M\"obius) generators correspond to the global $\mathfrak{sl}(2,\mathds{R})$ charges,
\begin{align} 
\mathcal{L}_0&=\oint \mathrm{d}z \ J^3(z)\ , &\mathcal{L}_{-1}&=\oint \mathrm{d}z \ J^+(z)\ , &\mathcal{L}_{1}&=\oint \mathrm{d}z \ J^-(z)\ .
\end{align}
\end{enumerate}
These requirements admit the following (symmetrical) solution \cite{Giveon:1998ns}
\be 
\mathcal{L}_m=\oint \mathrm{d}z \left((1-m^2) \gamma^m J^3+\frac{m(m-1)}{2} \gamma^{m+1} J^++\frac{m(m+1)}{2} \gamma^{m-1} J^- \right)(z)\ . \label{eq:spacetime Virasoro algebra bos}
\ee
By construction, these operators then map physical states onto physical states. One can directly compute their algebra as
\be \label{VirDDF}
[\mathcal{L}_m,\mathcal{L}_n]=(m-n)\, \mathcal{L}_{m+n}+\tfrac{k}{2}\, \mathcal{I}\, m(m^2-1)\, \delta_{m+n,0}\ ,
\ee
where
\be \label{Idef}
\mathcal{I}=\oint \mathrm{d}z\ \big(\gamma^{-1} \partial \gamma\big)(z)\ .
\ee
We will discuss the meaning of $\mathcal{I}$ below. For now, we remark that $\mathcal{I}$ commutes with all Virasoro generators $\mathcal{L}_m$ and is hence a central element of the algebra. One way to see this is to note that 
\begin{align}
[\mathcal{L}_m,\mathcal{I}]&= \Bigg(\hspace*{-.8cm}\oint\limits_{\hspace*{1.2cm}|z|>|w|}\hspace*{-.8cm} \mathrm{d}z\oint \mathrm{d}w -\hspace*{-.8cm}\oint\limits_{\hspace*{1.2cm}|z|<|w|}\hspace*{-.8cm} \mathrm{d}z\oint \mathrm{d}w \Bigg) \, \big(\gamma^{-1} \partial \gamma\big)(w)  \\
& \qquad \qquad \Big((1-m^2) \gamma^m J^3+\frac{m(m-1)}{2} \gamma^{m+1} J^
+\frac{m(m+1)}{2} \gamma^{m-1} J^- \Big)(z)  \nonumber \\[4pt]
&= \oint_0 \mathrm{d}w \oint_w \mathrm{d}z\ \left(-\frac{\gamma^m(w)}{(z-w)^2}-\frac{m (\gamma^{m-1} \partial \gamma)(w)}{z-w}\right) \label{OPE}\\
&=-m \oint_0 \mathrm{d}w\ \big(\gamma^{m-1} \partial \gamma\big)(w)\\
&=-\oint_0 \mathrm{d}w\ \partial \big(\gamma^m\big)(w)=0\ .
\end{align}
where the OPE in (\ref{OPE}) follows directly by inserting the Wakimoto representation \eqref{Wakib a}--\eqref{Wakib c} and using the free field OPEs \eqref{eq:betagamma OPE} and \eqref{eq:PhiPhi OPE}.

\subsubsection{Kac-Moody algebras}

In the superconformal situation to be discussed below, the internal manifold $X$ will contain an ${\rm S}^3$ factor, which can be described by an $\mathfrak{su}(2)$ WZW model. Whenever the world-sheet theory contains a WZW factor, we have an affine Kac-Moody algebra $\mathfrak{g}_{k_{\mathrm{G}}}$ on the world-sheet whose generators we denote by 
\be 
[K_m^a,K_n^b]=k_{\mathrm{G}}\, m \delta^{ab}\delta_{m+n,0}+i \tensor{f}{^{ab}_c} K^c_{m+n}\ ,
\ee
where $\tensor{f}{^{ab}_c}$ are the structure constants of the Lie algebra $\mathfrak{g}$ corresponding to the group $\mathrm{G}$. We may then construct DDF operators realising the corresponding symmetry in spacetime via \cite{Giveon:1998ns}
\be \label{KDDF}
\mathcal{K}^a_m= \oint \mathrm{d}z\ \big(\gamma^m K^a\big)(z)\ .
\ee
These operators are BRST invariant and lead to a Kac-Moody algebra in spacetime with commutation relations
\be \label{calK}
[\mathcal{K}^a_m,\mathcal{K}^b_n]=k_{\mathrm{G}}\, \mathcal{I} \, m \delta^{ab} \delta_{m+n,0}+i \tensor{f}{^{ab}_c} \mathcal{K}^c_{m+n}\ .
\ee
Here $\mathcal{I}$ is again defined by (\ref{Idef}).

\subsubsection{Higher Spins}

One may expect that a similar construction should also work for the other chiral world-sheet fields arising from $X$. In particular, we should be able to construct DDF operators associated to the (internal) Virasoro algebra associated to $X$, as well as for any higher spin generator of the chiral algebra associated to $X$. 
(This just reflects the fact that any world-sheet symmetry encoded by the presence of a chiral field should lead to a corresponding spacetime symmetry.)
These expectations are indeed borne out, and the relevant constructions are described in Appendix~\ref{app:higher spins}; since this does not seem to have been discussed in the literature before, we give a fairly detailed account of it there.

\subsection{The identity operator} \label{subsec:identity operator}

For the following it will be important to understand the structure of the central extension $\mathcal{I}$ of eq.~(\ref{Idef}). As we have seen above, ${\cal I}$ commutes with the Virasoro generators (and therefore also with all the other DDF operators), and hence acts as a constant in a given representation of the spacetime algebra.  Since its definition only involves $\gamma$, its value can only depend on the given $\mathfrak{sl}(2,\mathds{R})_k$ representation. To determine the relevant constants, we recall that the highest weight states of the $\mathfrak{sl}(2,\mathds{R})$ WZW model can be described by $|j,m\rangle$, see eq.~\eqref{jmfree}, 
in the unflowed sector. Since $|j,m\rangle$ does not contain $\beta$ (and $\gamma$ has regular OPE with itself as well as with $\partial\Phi$), it follows directly that $\mathcal{I}$ has trivial action in the unflowed sector 
\be \label{Ievalun}
\mathcal{I}\, |j,m \rangle=0\ .
\ee
On the other hand, upon spectral flow
\be 
\sigma^w(\mathcal{I})=\oint \mathrm{d}z\  \big(\gamma^{-1} z^{-w} \partial (\gamma z^w)\big)(z)=\mathcal{I}+w \, \mathds{1}\ .
\ee
Thus, we conclude that $\mathcal{I}$ acts as $w$ times the identity in the $w$-th spectrally flowed sector \cite{Giveon:1998ns}.

\subsection{The moding of the spacetime algebra} \label{subsec:locality and moding}

Next we want to analyse the structure of the algebra of DDF operators, in particular, that of the central terms. For this it will be important to analyse carefully the conditions under which the action of these DDF operators is well-defined. 

Let us first consider the unflowed sector. Because of (\ref{Ievalun}) together with the explicit formula for (\ref{Idef}), it follows that $\log(\gamma)(z)$ can be consistently defined without branch cut by\footnote{Strictly speaking, this argument only shows that $\log(\gamma)(z) $ can be defined without branch cut in a neighborhood of 0. However, this is all what is needed in the following.}
\be 
\log(\gamma)(z)\equiv \int_1^z \mathrm{d}w\ \big(\gamma^{-1} \partial \gamma\big)(w)\ .
\ee
Thus we conclude that 
\be \label{gammam}
\gamma^m(z)=\exp(m\log(\gamma)(z))
\ee
is a single-valued field for any (real) number $m \in \mathds{R}$. We should mention that all these expressions are well-defined; as $\gamma(z)$ has a regular OPE with itself, there is no normal-ordering ambiguity.

Next we consider the $w$-th spectrally flowed sector. Because of \eqref{wgam bos}, the $m$-th power of $\gamma$ becomes 
\be
\big(\sigma^w(\gamma)\big)^m(z)  = z^{-mw}  \gamma^m(z) \ . 
\ee
This therefore defines a single-valued field provided that $m \in \tfrac{1}{w} \mathds{Z}$. Since it is $\gamma^m$ that  appears in the definition of the various DDF operators, see eqs.~(\ref{eq:spacetime Virasoro algebra bos}), (\ref{KDDF}), as well as (\ref{WsDDF}), we conclude that we may take the mode numbers of the DDF operators to be fractional, with the fractional part being determined by the spectral flow,\footnote{In the unflowed sector we may take $n \in \mathds{R}$. We should remind the reader that the only physical states (except for the tachyon) that arise from the unflowed sector come from discrete representations.}
\be\label{wbos}
\hbox{$w$-th spectrally flowed sector:} \qquad 
n \ \in \tfrac{1}{w} \mathds{Z} \ .
\ee
We should stress that these fractionally moded operators map in general different $\mathfrak{sl}(2,\mathds{R})$ representations into one another. In particular, an oscillator $\mathcal{Z}_n$ with mode number $n$ carries charge $-n$ under $J^3_0$, and hence $\mathcal{Z}_n$ acts on the continuous representations as
\be\label{2.31}
\mathcal{Z}_n : \mathscr{C}^j_\lambda \rightarrow \mathscr{C}^j_{\lambda-n} \ . 
\ee
Thus for $n\not\in \mathds{Z}$, the two representations are inequivalent, but since both are part of the world-sheet  spectrum, these operators are still well-defined. Note that for the diagonal world-sheet spectrum (that is appropriate for the description of ${\rm AdS}_3$), the left- and right-moving values of $\lambda$ agree; this turns out to incorporate the orbifold projection in the $w$-cycle twisted sector, see \cite{Giribet:2018ada,Gaberdiel:2018rqv,Eberhardt:2018ouy} and the comments below in Section~\ref{sec:spacetime action}.

We should also mention that on the discrete representations we necessarily need $n \in \mathds{Z}$ (since otherwise the image lies in a representation that is not part of the world-sheet spectrum). In the following we shall therefore only consider the continuous world-sheet representations; we will comment on the role of the discrete world-sheet representations in Section~\ref{subsec:discretereps}.

\subsubsection{Untwisting} \label{subsec:untwisting}

As we have just seen, the spacetime generators are naturally fractionally moded when acting in  the $w$-th spectrally flowed continuous representations, see eq.~(\ref{wbos}). This suggests that these worldsheet representations give rise to the $w$-cycle twisted sector of a symmetric product orbifold from the viewpoint of the spacetime CFT, see also \cite{Giribet:2018ada,Gaberdiel:2018rqv,Eberhardt:2018ouy}. In order to read off the structure of the underlying seed theory we can `untwist' these generators. Let us explain this for the case of the (overall) Virasoro generators. We propose to define the 
untwisted generators $\widehat{\mathcal{L}}_m$ via
\be 
\mathcal{L}_{\frac{m}{w}}=\frac{1}{w}\widehat{\mathcal{L}}_m+\frac{k\, (w^2-1)}{4w}\, \delta_{m,0}\ , \label{eq:Virasoro algebra untwisted}
\ee
where $m$ now takes values in $\mathds{Z}$. While the central term for the original $\mathcal{L}_n$ modes depends on $w$ (because $\mathcal{I}=w\, \mathds{1}$), the above correction term ensures that the $\widehat{\mathcal{L}}_m$  satisfy a Virasoro algebra with $c=6k$, independently of $w$. Indeed, it follows from (\ref{VirDDF}) that $[\widehat{\mathcal{L}}_m,\widehat{\mathcal{L}}_{-m}]$ equals
\begin{align}
[\widehat{\mathcal{L}}_m,\widehat{\mathcal{L}}_{-m}] & = 2 m  w \,  \mathcal{L}_0 + 
\frac{k}{2} w^2 \frac{m}{w} \Bigl( \frac{m^2}{w^2} - 1\Bigr) w  \\
& = 2 m \, \widehat{\mathcal{L}}_0 + m\, \frac{k}{2}  \Bigl[ (w^2-1)  +  (m^2-w^2) \Bigr]  \\ 
& = 2 m \, \widehat{\mathcal{L}}_0 + m\, (m^2-1) \, \frac{k}{2}  \ . 
\end{align}
The relation between the two sets of modes in eq.~(\ref{eq:Virasoro algebra untwisted}) has exactly the same form as for the case of a symmetric orbifold, where the $\mathcal{L}_n$ modes act on the covering space, while the $\widehat{\mathcal{L}}_m$ generators are those of the seed theory, see e.g.\  \cite{Lunin:2000yv, Burrington:2018upk, Roumpedakis:2018tdb}. 

The analysis for the other DDF operators is similar. For example, for the current algebra (\ref{calK}), the untwisted generators are defined via
\be\label{Ktrans}
\mathcal{K}^a_{\frac{m}{w}}= \widehat{\mathcal{K}}^a_m \ , 
\ee
leading to an affine Kac-Moody algebra of level $k_{\mathrm{G}}$ for the seed theory. This formula generalises directly also to the higher spin DDF operators of Appendix~\ref{app:higher spins}: for a primary field of spin $s$, the untwisted generators are defined via
\be\label{Wtrans}
\mathcal{W}_{\frac{m}{w}}^{(s)}=w^{1-s}\, \widehat{\mathcal{W}}^{(s)}_m \ , 
\ee
and these untwisted generators then have exactly the same commutation relations as in the original world-sheet chiral algebra. Note that (\ref{Ktrans}) and (\ref{Wtrans}) are simply the transformation rules  of a primary field of conformal weight $s$ under the map $z \mapsto z^w$, which relates the modes defined on the covering space to those of the base space. The correction term in (\ref{eq:Virasoro algebra untwisted}) then reflects the fact that the stress-energy tensor is only quasiprimary, and that its transformation rule therefore also involves the Schwarzian derivative.

\subsubsection{The seed theory}

The above considerations suggest that the $w=1$ sector corresponds to the untwisted sector of a symmetric product orbifold in spacetime. As a consequence, the DDF operators in the $w=1$ sector describe the chiral algebra of the seed theory of the symmetric product. The construction we have discussed so far shows that this chiral algebra contains the chiral algebra of $X$, together with the overall Virasoro tensor of central charge $c=6k$, under which the primary fields of $X$ transform also as primary fields. In order to elucidate the structure of this algebra, it is convenient to decouple the $X$ factor by defining the (coset) Virasoro tensor
\be 
\mathcal{L}_m^\text{L}=\mathcal{L}_m-\mathcal{L}_m^\text{m}\ , \label{eq:bosonic decoupling Virasoro}
\ee
where we have subtracted the `matter' Virasoro algebra $\mathcal{L}^\text{m}$ associated to $X$, see Appendix~\ref{app:higher spins} for its explicit construction from the worldsheet. The modes of $\mathcal{L}_m^\text{L}$ commute by construction with all modes of $X$ and lead to a Virosoro algebra of central charge \cite{Seiberg:1999xz}
\be 
c^\text{L}=6k-c_X=6k-\left(26-\frac{3k}{k-2}\right)=1+\frac{6(k-3)^2}{k-2}\ . \label{eq:bosonic Liouville central charge}
\ee
Here, we have used that the string background is critical, see eq.~\eqref{eq:criticality}. Thus, the spacetime algebra is a Virasoro algebra of central charge $c^\text{L}$ together with the (decoupled) chiral algebra of the internal CFT. We note that the Virasoro algebra can be represented by a Liouville field with 
\be\label{Qpdef}
Q' = \frac{k-3}{\sqrt{k-2}} \ . 
\ee
This will play an important role in the following. 

\subsection{Identifying Liouville theory on the world-sheet}\label{sec:Identify}

The previous discussion now suggests that the seed theory of the spacetime symmetric orbifold is given by
\be \label{spacetimeseed}
\left[\text{Liouville with }c^\text{L}=1+\frac{6(k-3)^2}{k-2}\right] \times X\ .
\ee
Since we have already shown that the spacetime theory has the corresponding symmetry generators, it remains to match the spectrum of the two theories. Recall that bosonic Liouville theory is believed to be uniquely characterised by having Virasoro symmetry, together with the full spectrum of primary fields \cite{Ribault:2014hia, Collier:2017shs}. 

To match the spectrum, we look at the string theory mass shell condition, which reads in the bosonic case
\be 
\frac{-j(j-1)}{k-2}-wh+\frac{k}{4} w^2+h_\text{int}+N=1\ ,
\ee
where $h$ is the conformal weight of the state in the dual CFT, $N$ the excitation number on the worldsheet and $h_\text{int}$ the conformal weight of the state in the internal CFT. For $w>0$, we can immediately solve for $h$,\footnote{This step only works in this manner for the continuous world-sheet representations since the $J^3_0$ eigenvalues (modulo one) are not already determined by $j$, but are parametrised by the independent parameter $\lambda$.} which gives
\be \label{hst}
h=- \frac{\alpha(\alpha-Q)}{w}+\frac{kw^2-4}{4w}+\frac{h_\text{int}+N}{w}\ ,
\ee
where we have rewritten the conformal dimension of the $\mathfrak{sl}(2,\mathds{R})_k$ part using eq.~(\ref{worldsheetLiouville}). Next we observe that the first two terms can be rewritten as 
\begin{align}
- \frac{\alpha(\alpha-Q)}{w} +\frac{kw^2-4}{4w} & = - \frac{1}{w}\Bigl(\alpha - \frac{Q}{2}\Bigr)^2 + \frac{Q^2}{4w} +\frac{kw^2-4}{4w} \\[3pt]
& = - \frac{1}{w}\Bigl(\alpha - \frac{Q}{2}\Bigr)^2 + \frac{{Q'}^2}{4w}+\frac{k}{4w}(w^2-1)\\[3pt]
& \equiv - \frac{\alpha'(\alpha'-Q')}{w}+\frac{c}{24w}(w^2-1)  \ , 
\end{align}
where we have used that (\ref{worldsheetLiouville}) and (\ref{Qpdef}) imply that 
\be
Q^2 = {Q'}^2 - (k-4) \ , 
\ee
and identified 
\be\label{alphap}
\alpha' - \frac{Q'}{2}  = \pm \Bigl(\alpha - \frac{Q}{2}\Bigr) \ . 
\ee
Moreover, we have used the fact that the seed theory has central charge $c=6k$.
Thus the spacetime theory has exactly the spectrum of the symmetric orbifold of (\ref{spacetimeseed}), where the last term 
$\frac{c}{24w}(w^2-1)$ is the ground state conformal weight in the $w$-cycle twisted sector. 
Furthermore, because of (\ref{alphap}), the continuous representations on the world-sheet (that correspond precisely to the usual representation of the Liouville field $\Phi$, see the discussion below eq.~(\ref{worldsheetLiouville})), map one-to-one to the usual Liouville representations of the spacetime $Q'$ theory. 

We have therefore shown that the continuous representations  on the world-sheet lead precisely to the symmetric orbifold (in the large $N$ limit)
\be \label{stsymorb}
\text{Sym}^N\left(\left[\text{Liouville with }c^\text{L}=1+\frac{6(k-3)^2}{k-2}\right] \times X\right)
\ee
in spacetime. 

Let us mention that for $k=3$, the Liouville part has $c^\text{L}=1$,\footnote{There are actually two theories with this spectrum for $c^\text{L}=1$, one being Liouville theory and the other being the Runkel-Watts theory \cite{Runkel:2001ng, Gaberdiel:2011aa}, see \cite{Schomerus:2003vv, McElgin:2007ak, Ribault:2015sxa}. From what we have shown, it is not entirely clear what the correct theory should be. However, Liouville theory exists for any real $k$, whereas non-analytic Liouville theory (of which the Runkel-Watts theory is a particular case) only exists for $b^2\in \mathds{Q}$, where $c=1+6(b+b^{-1})^2$. Since we expect a continuous behaviour in $k$, it is natural that the correct theory should always be Liouville theory, also at $c^\text{L}=1$.}
and in particular there is no gap in the spectrum. This ties in with the observation made in \cite{Gaberdiel:2017oqg},  that there are massless higher spin fields appearing for this special amount of flux. Contrary to what happens in the supersymmetric setting \cite{Eberhardt:2018ouy}, this does, however, not mean that the long string continuum disappears. We should mention that the continuum for $k=3$ is somewhat reminiscent of the light states spectrum of higher spin theories, see e.g.\ \cite{Gaberdiel:2013cca}.

\subsection{Discrete representations}\label{subsec:discretereps}

We end this bosonic analysis with a brief discussion of the role of the discrete world-sheet representations. As we have seen above, the continuous world-sheet representations lead precisely to the spacetime spectrum of the symmetric orbifold (\ref{stsymorb}), which defines a well-defined spacetime CFT by itself. One may therefore wonder what role the states from the world-sheet discrete representations should play?

It follows from the analysis in Section~\ref{sec:Identify} that the discrete representations can give rise to physical states that lie below the Liouville gap, see in particular eq.~(\ref{alphap}). However, the analysis of the discrete representations is somewhat complicated since the $J^3_0$ eigenvalue, modulo integers, equals the spin $j$, and hence one cannot just solve the mass-shell condition as in eq.~(\ref{hst}). As a consequence, the existence of a physical state also depends on the precise value of $h_{\rm int}$ (and $N$). Furthermore, the Maldacena-Ooguri bound \cite{Maldacena:2000hw} constrains $j$ to lie in the interval $(\tfrac{1}{2},\tfrac{k-1}{2})$.

\begin{figure}
\includegraphics[width=\textwidth]{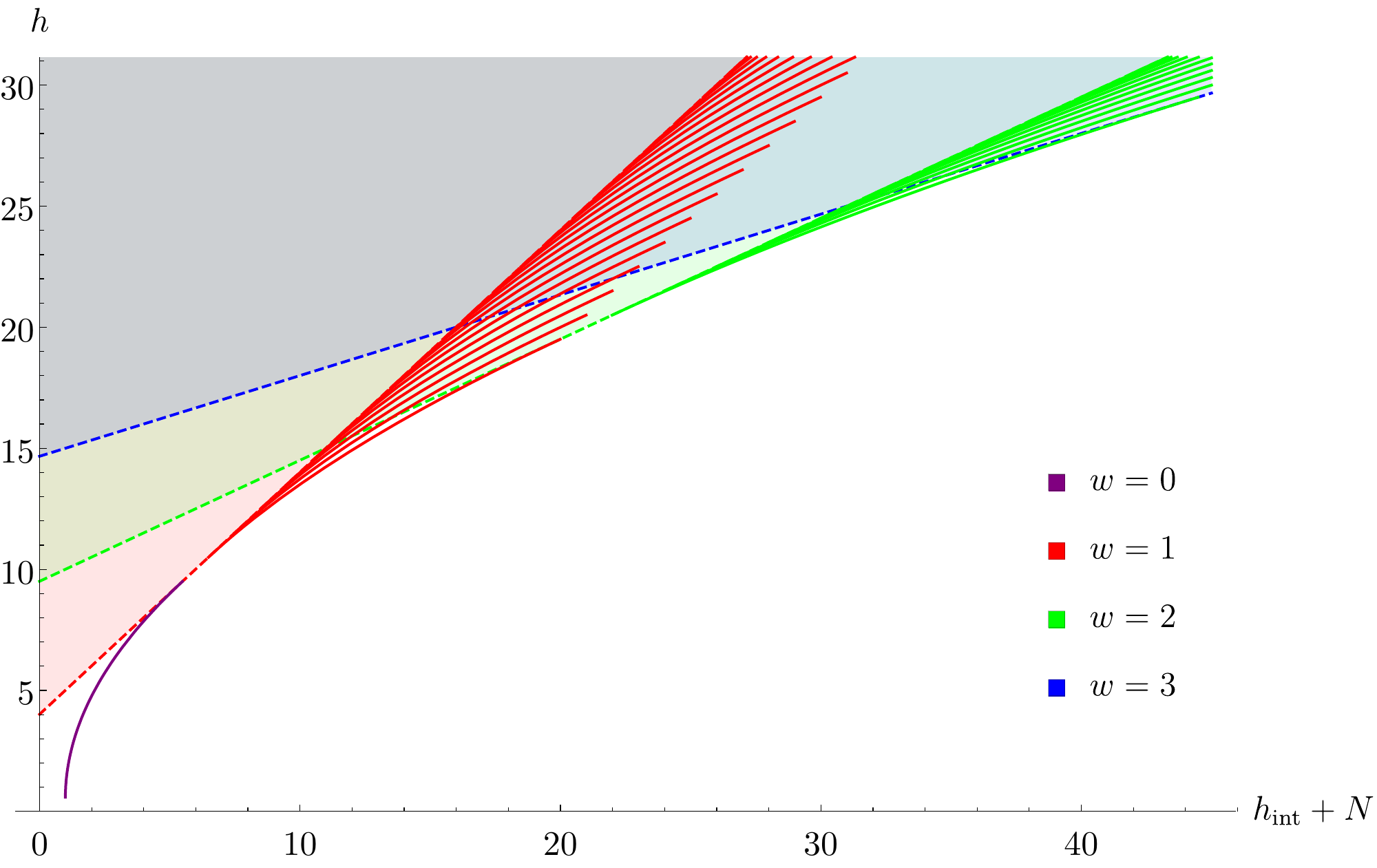}
\caption{The spacetime conformal weight in dependence of the internal conformal weight $h_{\rm int}+N$. For definiteness, we have chosen $k=20$. Dashed lines indicate the bottom of the continuum in the respective sector. We have plotted the discrete representation as solid lines. There are several lines which correspond to the choice of state in a particular $\mathfrak{sl}(2,\mathds{R})$ representation.} \label{fig:discrete}
\end{figure}

We have analysed systematically which spacetime states arise from the discrete representations (as a function of $h_{\rm int}+ N$), and the picture that emerges is the following,\footnote{This also ties in with the findings of  \cite{Ferreira:2017pgt}.} see Fig.~\ref{fig:discrete}: the discrete representations with spectral flow $w$ lead to spacetime states that either lie inside the continuum --- this is the case for the vast majority of these states --- or are a number of isolated states just below the Liouville gap. (For large $k$, the distance to the Liouville gap scales as $k$.) These isolated states in turn lie above a line that interpolates between the gaps coming from the $w$'th and $w+1$'st twisted sector of the symmetric orbifold. The fact that these states lie below the symmetrically orbifolded Liouville gap will be important below in the susy setting, since the discrete representations  account for the BPS states of the spactime theory (which are not part of the symmetric orbifold of ${\cal N}=4$ Liouville theory). However, apart from these special cases, the resulting spacetime states do not seem to correspond to special Liouville representations --- in fact, given that they depend sensitively on the value of $h_{\rm int}+ N$, this cannot be otherwise. 

As explained in \cite{Maldacena:2001km}, the discrete representations on the world-sheet lead to non-normalisable operators in the spacetime CFT that are not really part of the spacetime CFT. In particular, not all of their correlation functions are well-defined; for example, in the unflowed sector the condition for the $n$-point correlator to be well-defined (and make physical sense) is that 
\be\label{bound}
\sum_i j_i < k + n -3 \ , 
\ee
see footnote~19 of \cite{Maldacena:2001km}. As we have seen above, see eq.~(\ref{alphap}), the dictionary to the spacetime CFT implies that the discrete representation with spin $j=\frac{\alpha}{Q}$, corresponds to a field in the spacetime Liouville theory with 
\be
\alpha' - \tfrac{Q'}{2} = \bigl(j-\tfrac{1}{2} \bigr) Q \ . 
\ee
Thus the above bound (\ref{bound}) becomes
\be
\sum_i (\alpha'_i - \tfrac{Q'}{2}) = \sum_i P_i  < \bigl( k + \tfrac{1}{2} (n-6)) Q  \cong Q' \ , 
\ee
where $P_i$ are the so-called Liouville momenta, and we have used that $(k-3) Q = Q'$. In the semiclassical limit, $k$ and hence $Q'$ are large, and this condition therefore becomes 
\be
\sum_i P_i  < Q' = b' + (b')^{-1} \cong b' \ , 
\ee
where $b'$ is the parameter that appears in the exponential of the Liouville potential. This has now a natural interpretation, namely that the Liouville potential needs to dominate over the excitations of the fields in the correlator. This therefore realises the idea of the quantum mechanical toy model in Section~3.2 of  \cite{Maldacena:2001km}, and thus nicely ties in with their findings.

\section{A review of superstrings on \texorpdfstring{$\boldsymbol{\mathrm{AdS}_3 \times \mathrm{S}^3 \times \mathbb{T}^4}$}{AdS3xS3xT4}}\label{sec:review of strings}

In the following sections we shall repeat the above analysis for the supersymmetric setting. We shall concentrate on the case of superstring theory on $\mathrm{AdS}_3 \times \mathrm{S}^3 \times \mathbb{T}^4$. There are two formalisms for describing the background, the RNS formalism \cite{Maldacena:2000hw} and the hybrid formalism \cite{Berkovits:1999im}. The RNS formalism is technically simpler, since it involves only bosonic WZW models and free fermions on the worldsheet. On the other hand, the hybrid formalism makes spacetime supersymmetry manifest. 
Moreover as demonstrated in \cite{Eberhardt:2018ouy}, it is well-defined for all values of NS-NS background flux $k$, in particular for $k=1$. In the following, we shall actually use a mixture of both formalisms, treating the fields of $\mathbb{T}^4$ in the RNS-formalism, and the fields of $\mathrm{AdS}_3 \times \mathrm{S}^3$ in the hybrid formalism.
\subsection{The RNS formalism} \label{subsec:RNS formalism}
In the RNS formalism, the worldsheet theory takes the form \cite{Giveon:1998ns, Maldacena:2000hw}
\be 
\mathfrak{sl}(2,\mathds{R})^{(1)}_k \oplus \mathfrak{su}(2)^{(1)}_k \oplus \big(\mathfrak{u}(1)^{(1)}\big)^4\ ,
\ee
where $\mathfrak{sl}(2,\mathds{R})^{(1)}_k$ describes the $\mathrm{AdS}_3$ factor, $\mathfrak{su}(2)^{(1)}_k$ the $\mathrm{S}^3$ factor and the flat torus directions are represented by four $\mathfrak{u}(1)$ currents. The superscript $(1)$ indicates that these are the corresponding $\mathcal{N}=1$ superconformal current algebras. We can decouple the free fermions from the affine generators and hence get 
\begin{align}
\mathfrak{sl}(2,\mathds{R})^{(1)}_k &\cong \mathfrak{sl}(2,\mathds{R})_{k+2} \oplus \text{3 free fermions}\ , \\
\mathfrak{su}(2)^{(1)}_k &\cong \mathfrak{su}(2)_{k-2} \oplus \text{3 free fermions}\ .
\end{align}
We shall denote the decoupled $\mathfrak{sl}(2,\mathds{R})_{k+2}$ currents by $\mathscr{J}^a$, the decoupled $\mathfrak{su}(2)_{k-2}$ currents by $\mathscr{K}^a$, and the free bosons of $\mathbb{T}^4$ by $\partial X^{\alpha}$ and $\partial \bar{X}^{\alpha}$. Here, $a=\pm$, 3 is an adjoint index of $\mathfrak{sl}(2,\mathds{R})$ or $\mathfrak{su}(2)$, respectively, while $\alpha=\pm$. Note that we have paired the four bosons of $\mathbb{T}^4$  into two complex bosons. The index $\alpha=\pm$ will turn out to be a spinor index of the outer automorphism group of small $\mathcal{N}=4$ supersymmetry.

Moreover, we have ten fermions on the worldsheet transforming in the adjoint representation of the bosonic groups. We denote the $\mathfrak{sl}(2,\mathds{R})$ fermions by $\psi^a$, and the $\mathfrak{su}(2)$ fermions by $\chi^a$, where again $a=\pm$, $3$. Finally, we pair the four fermions of $\mathbb{T}^4$ together into two complex fermions, which we denote by $\lambda^\alpha$ and $\bar{\lambda}^\alpha$. $\lambda^\alpha$ is the $\mathcal{N}=1$ superpartner of the bosons $\partial X^\alpha$ on the worldsheet, and similarly for $\bar{\lambda}^\alpha$ and $\partial \bar{X}^{\alpha}$. The $\alpha$-index is again a spinor index of the outer automorphism $\mathfrak{su}(2)$ of the small $\mathcal{N}=4$ spacetime supersymmetry. We will see below that the fermions $\lambda^\alpha$ and $\bar{\lambda}^\alpha$ give rise to four fermions in spacetime which transform as doublets under the $\mathcal{N}=4$ R-symmetry. The (anti)commutation relation of the modes of these fields are summarised in Appendix~\ref{app:OPEs}.

As before we want to quantise this theory via BRST quantisation. For this, we need the $\mathcal{N}=1$ superconformal structure on the worldsheet, which is explicitly given by
\begin{align}
T(z)&=\frac{1}{k}\Big(-\mathscr{J}^3\mathscr{J}^3+\tfrac{1}{2}\big(\mathscr{J}^+\mathscr{J}^-+\mathscr{J}^-\mathscr{J}^+\big)+\psi^3 \partial \psi^3-\tfrac{1}{2}\big(\psi^+\partial \psi^-+\psi^-\partial \psi^+\big)\Big) \nonumber\\
&\qquad+\frac{1}{k}\Big(\mathscr{K}^3\mathscr{K}^3+\tfrac{1}{2}\big(\mathscr{K}^+\mathscr{K}^-+\mathscr{K}^-\mathscr{K}^+\big)-\chi^{3}\partial \chi^{3}-\tfrac{1}{2}\big(\chi^{+}\partial\chi^{-}+\chi^{-}\partial \chi^{+}\big)\Big)\nonumber\\
&\qquad+\epsilon_{\alpha\beta}(\partial X^\alpha \partial \bar{X}^\beta)+\frac{1}{2}\epsilon_{\alpha\beta}\big(\partial \lambda^\alpha\bar{\lambda}^\beta-\lambda^\alpha\partial \bar{\lambda}^\beta\big)\ , \label{eq:worldsheet Virasoro}\\
G(z)&=-\frac{1}{k}\Big(-\mathscr{J}^3\psi^3+\tfrac{1}{2}\big(\mathscr{J}^+\psi^-+\mathscr{J}^-\psi^+\big)-\tfrac{1}{k}(\psi^3\psi^+\psi^-)\Big) \nonumber\\
&\qquad-\frac{1}{k}\Big(\mathscr{K}^3\chi^3+\tfrac{1}{2}\big(\mathscr{K}^+\chi^-+\mathscr{K}^-\chi^+\big)+\tfrac{1}{k}(\chi^3\chi^+\chi^-)\Big) \nonumber\\
&\qquad+\frac{1}{2} \epsilon_{\alpha\beta} \big(\partial X^\alpha \bar{\lambda}^\beta-\partial \bar{X}^\alpha\lambda^\beta\big)\ , \label{eq:worldsheet supercurrent} 
\end{align}
where here, as in the following, normal-ordering is always understood. 
Moreover, we introduce the standard ghosts of the superstring, i.e.\ a $bc$ system with $\lambda =2$, and a $\beta\gamma$ system with $\lambda=\frac{3}{2}$, satisfying (see Appendix~\ref{app:freeconv} for our conventions)\footnote{In order to distinguish the superconformal ghosts from the $\beta\gamma$ system that appears in the Wakimoto realisation of $\mathfrak{sl}(2,\mathds{R})$, we denote them with a hat.}
\be 
b(z)c(w) \sim \frac{1}{z-w}\ , \qquad \hat{\beta}(z) \hat{\gamma}(w) \sim -\frac{1}{z-w}\ .
\ee
These fields also generate an $\mathcal{N}=1$ superconformal structure with
\begin{align}
T_\text{gh}(z)&=-2b (\partial c) - (\partial b) c-\tfrac{3}{2}\hat{\beta} (\partial \hat{\gamma}) -\tfrac{1}{2} (\partial \hat{\beta}) \hat{\gamma}\ , \\
G_\text{gh}(z)&=(\partial \hat{\beta}) c+\tfrac{3}{2}\hat{\beta} (\partial c) -\tfrac{1}{2} b \hat{\gamma}\ .
\end{align}
The standard BRST operator of the superstring is then given by
\be 
Q=\oint \mathrm{d}z \Big(c\big(T+\tfrac{1}{2}T_\text{gh}\big)+ \hat{\gamma}\big(G+\tfrac{1}{2}G_\text{gh}\big)\Big)\ .
\ee
We split $Q$ into three pieces according to their $\hat{\beta}\hat{\gamma}$ ghost number \cite{Friedan:1985ge}
\be
Q=Q_0+Q_1+Q_2\ , \label{eq:BRST charge}
\ee
where
\begin{align}
Q_0&=\oint \mathrm{d}z\ c \big(T+T_\text{gh}\big)+b(\partial c)c + \tfrac{3}{4} \partial \bigl( \hat{\gamma} \hat{\beta} c \bigr)  = 
\oint \mathrm{d}z\ c \big(T+T_\text{gh}\big)+b(\partial c)c 
\ , \label{eq:BRST charge Q0}\\
Q_1&=\oint \mathrm{d}z\ \hat{\gamma} G\ , \label{eq:BRST charge Q1}\\
Q_2&=-\tfrac{1}{4} \oint \mathrm{d}z\  b \hat{\gamma} \hat{\gamma}\ . \label{eq:BRST charge Q2}
\end{align}
$Q_0$ is the BRST operator of the bosonic string, for which the $\hat{\beta}\hat{\gamma}$ ghosts are treated as additional matter fields. BRST invariance under $Q_2$ is usually trivially (at least in the canonical picture), and the important piece of the BRST charge is $Q_1$.

In the following, we will always use the bosonised form of the superconformal ghosts, i.e.\ we introduce free bosons with background charge $Q_\phi=2$ and $Q_\chi=-1$ and OPEs 
\begin{align}
\phi(z)\phi(w)&\sim -\log(z-w)\ , & \chi(z)\chi(w)&\sim\log(z-w)\ ,
\end{align}
and write 
\begin{align} 
\hat{\beta}&= \mathrm{e}^{-\phi} \mathrm{e}^{\chi} \partial \chi\ , 
& \hat{\gamma}&=\mathrm{e}^{-\chi} \mathrm{e}^{\phi}\ . 
\end{align}
The $\phi$-boson then has $c=13$, while the $\chi$-boson yields $c=-2$, see Appendix~\ref{app:freeconv} for our conventions. (This then reproduces the central charge $c=11$ of the $\hat{\beta}\hat{\gamma}$ system.) We also define the $(\xi,\eta)$ pair via 
\be
\xi = \mathrm{e}^{\chi} \ , \qquad \eta = \mathrm{e}^{-\chi} \ , 
\ee
where $h(\xi)=0$ and $h(\eta) = 1$. Finally, the picture raising operation on vertex operators is defined by $Z=-2[Q,\xi\, \boldsymbol{\cdot}\, ]$.

\subsection{The hybrid formalism} \label{subsec:hybrid formalism}

In order to rewrite these degrees of freedom in terms of the hybrid formalism, we now bosonise the ten fermions, i.e.\ we introduce bosons via 
\begin{subequations}
\begin{align}
\partial H_1&=\frac{1}{k}(\psi^+\psi^-)\ ,  & \partial H_2 &=\frac{1}{k}(\chi^+\chi^-)\ , & \partial H_3&=\frac{2}{k} (\psi^3 \chi^3)\ , \label{eq:bosonisation a}\\
\partial H_4&=(\lambda^+ \bar{\lambda}^-)\ , & \partial H_5&=-(\lambda^- \bar{\lambda}^+)\ .\label{eq:bosonisation b}
\end{align}
\end{subequations}
These bosons satisfy the standard OPEs (with vanishing background charge $Q_H=0$)
\be 
H_i(z) H_j(w) \sim \delta_{ij}\log(z-w)\ ,
\ee
and we can express the fermions in terms of them as 
\begin{subequations}
\begin{align}
\psi^\pm&=\sqrt{k}\mathrm{e}^{\pm H_1}\ , & \chi^\pm&=\sqrt{k}\mathrm{e}^{\pm H_2} &  \psi^3 \mp \chi^3&=\sqrt{k} \mathrm{e}^{\pm H_3}\ , \\
\lambda^+&=\mathrm{e}^{H_4} , & \lambda^-&= \mathrm{e}^{-H_5}\ , & \bar{\lambda}^+&=\mathrm{e}^{H_5} , & \bar{\lambda}^-&=\mathrm{e}^{-H_4}\ .
\end{align}
\end{subequations}
Here and in the following we will suppress cocycle factors. The final step consists of refermionising these bosons, i.e.\ by considering the fermionic generators that can be constructed out of these bosons as 
\begin{subequations}
\begin{align}
p^{\alpha\beta}&=\mathrm{e}^{\frac{\alpha}{2} H_1+\frac{\beta}{2}H_2+\frac{\alpha\beta}{2}H_3+\frac{1}{2}H_4+\frac{1}{2}H_5-\frac{1}{2}\phi}\ , \label{pdef}\\
\theta^{\alpha\beta}&=\mathrm{e}^{\frac{\alpha}{2} H_1+\frac{\beta}{2}H_2-\frac{\alpha\beta}{2}H_3-\frac{1}{2}H_4-\frac{1}{2}H_5+\frac{1}{2}\phi}\ ,  \label{thetadef}\\
\Psi^+&=\mathrm{e}^{H_4-\phi+\chi}\ , \\
\Psi^-&=\mathrm{e}^{-H_5-\phi+\chi}\ , \\
\bar{\Psi}^+&=\mathrm{e}^{H_5+\phi-\chi}\ , \\
\bar{\Psi}^-&=\mathrm{e}^{-H_4+\phi-\chi}\ .
\end{align}
\end{subequations}
These fields constitute again a collection of fermionic first-order systems
\begin{align}\label{fermfirst}
p^{\alpha\beta}(z)\theta^{\gamma\delta}(w)&\sim \frac{\epsilon^{\alpha\gamma}\epsilon^{\beta\delta}}{z-w}\ , \\
\Psi^\alpha(z)\bar{\Psi}^\beta(w)&\sim \frac{\epsilon^{\alpha\beta}}{z-w}\ , \label{fermsec}
\end{align}
where $p^{\alpha\beta}$ and $\Psi^\alpha$ have conformal dimension equal to one, while $\theta^{\alpha\beta}$ and $\bar{\Psi}^\alpha$ have conformal dimension equal to zero. 
In fact the four fields $p^{\alpha\beta}$ describe four of the eight spacetime supercharges in the canonical ghost picture, so the hybrid formalism makes half of spacetime supersymmetry manifest. We will see in the next subsection that the other four supercharges are also (almost) manifest.

The fermionic first-order systems (\ref{fermfirst}) and (\ref{fermsec}) describes six $bc$ pairs each with $\lambda=1$, thus giving rise altogether to $c=-12$. On the other hand, we started with ten fermions (giving $c=5$), as well as the $\hat{\beta}\hat{\gamma}$ superconformal ghosts with $c=11$. Thus we are missing central charge $c=28$, which is accounted for by the boson 
\be \label{rhoboson}
\rho=2\phi-H_4-H_5-\chi\ ,
\ee
which has background charge $Q=3$ in the conventions of Appendix~\ref{app:freeconv}, and will serve as a bosonised ghost in the hybrid formalism. Thus we have rewritten the fermionic degrees of freedom of the NS-R formalism in terms of the fermionic first-order system (\ref{fermfirst}) and (\ref{fermsec}), as well as the boson (\ref{rhoboson}). 

\subsection{Supergroup generators}

The final step consists of assembling the (unchanged) bosonic fields $\mathscr{J}^a$, $\mathscr{K}^a$ together with the fermions $p^{\alpha\beta}$ and $\theta^{\alpha\beta}$ into the current algebra for the superalgebra $\mathfrak{psu}(1,1|2)_k$; in fact, this is just the Wakimoto representation for this superalgebra. More specifically, we define
\begin{subequations}
\begin{align}
J^{\text{(f)}a}&=\tfrac{1}{2}c_a (\sigma^a)_{\alpha\mu} \epsilon_{\beta\nu} (p^{\alpha\beta} \theta^{\mu\nu})\ , \\
K^{\text{(f)}a}&=\tfrac{1}{2}\epsilon_{\alpha\mu} (\sigma^a)_{\beta\nu} (p^{\alpha\beta} \theta^{\mu\nu})\ ,
\end{align}
\end{subequations}
which generate the $\mathfrak{sl}(2,\mathds{R})_{-2} \oplus \mathfrak{su}(2)_2$ algebra. The full $\mathfrak{psu}(1,1|2)_k$-generators are then given as 
\begin{subequations}
\begin{align}
J^a&=\mathscr{J}^a+J^{\text{(f)}a}\ , \label{eq:Wakimoto representation J}\\
K^a&=\mathscr{K}^a+K^{\text{(f)}a}\ , \label{eq:Wakimoto representation K}\\
S^{\alpha\beta+}&=p^{\alpha\beta}\ , \label{eq:Wakimoto representation S+}\\
S^{\alpha\beta-}&=k \partial \theta^{\alpha\beta}+c_a \tensor{(\sigma_a)}{^\alpha_\gamma} \big(\mathscr{J}^a+\tfrac{1}{2}J^{\text{(f)}a}\big)\theta^{\gamma\beta}-\tensor{(\sigma_a)}{^\beta_\gamma} \big(\mathscr{K}^a+\tfrac{1}{2}K^{\text{(f)}a}\big)\theta^{\alpha\gamma}\ . \label{eq:Wakimoto representation S-}
\end{align}
\end{subequations}
One checks by a direct calculation, see also \cite{Bars:1990hx, Berkovits:1999im,Gotz:2006qp}, that these generators then satisfy the relations of $\mathfrak{psu}(1,1|2)_k$ that are spelled out in Appendix~\ref{subapp:psu112k WZW-model}. Moreover, we have used the conventions of \eqref{eq:sigma matrices conventions udd} for the sigma-matrices.
\medskip

Thus we conclude that the worldsheet theory in the hybrid formalism is generated by 
\be 
\mathfrak{psu}(1,1|2)_k \oplus \text{topologically twisted }\mathbb{T}^4\oplus\text{ ghosts}\ .
\ee
Here the topologically twisted $\mathbb{T}^4$ is described by the bosons $\partial X^\alpha$, $\partial \bar{X}^\alpha$, together with the (topologically twisted) fermions $\Psi^\alpha$ and $\bar{\Psi}^\alpha$, while the 
ghosts consist of the bosonic $(b,c)$ ghosts together with the $\rho$ ghost. In order for this to make sense one must also be able to rewrite the BRST operator (as well as the $(\xi,\eta)$ pair) in terms of these redefined fields, and this is indeed possible, see \cite{Berkovits:1999im} for details.

As shown in \cite{Eberhardt:2018ouy}, the WZW model $\mathfrak{psu}(1,1|2)_1$ can be consistently defined and hence the hybrid formalism provides a good definition of the $k=1$ worldsheet theory. This is one of the main motivations for us to use it here.

\section{The \texorpdfstring{$\boldsymbol{\mathfrak{psu}(1,1|2)_k}$}{psu(1,1|2)k} WZW model} \label{sec:psu112k WZW model}

In the next step we discuss the $\mathfrak{psu}(1,1|2)_k$ WZW-model and its vertex operators. Details about the OPEs are collected in Appendix~\ref{app:OPEs}.

\subsection{Wakimoto representation of \texorpdfstring{$\mathfrak{sl}(2,\mathds{R})_{k+2}$}{sl(2,R)(k+2)} and vertex operators} \label{subsec:sl2R Wakimoto representation}

We recall from Section~\ref{subsec:vertex operatorsb} that 
\be 
|j,m \rangle=\oint \mathrm{d}z\ \gamma^{-j-m} \mathrm{e}^{\, j\sqrt{\frac{2}{k}}\Phi}(z)|0\rangle \label{eq:sl2R highest weight states}
\ee
defines an affine highest weight state of an $\mathfrak{sl}(2,\mathds{R})_{k+2}$ representation.\footnote{Note that  the decoupled $\mathfrak{sl}(2,\mathds{R})_{k+2}$ algebra is now at level $k+2$, and hence $k$ is shifted by $+2$ relative to the formulae of Section~\ref{sec:bosonic}.}
These states can also be viewed as part of a $\mathfrak{psu}(1,1|2)$ multiplet. Since in the Wakimoto representation $S^{\alpha\beta+}=p^{\alpha\beta}$, these states are annihilated by all supercharge zero modes of that form. In order to match the conventions of \cite{Eberhardt:2018ouy}, we now shift $j$ by one half, and define 
\be 
|j,m,0,\uparrow\rangle=\oint \mathrm{d}z\ \gamma^{-j-m+\frac{1}{2}} \mathrm{e}^{(j-\frac{1}{2})\sqrt{\frac{2}{k}}\Phi}(z)|0\rangle\ , \qquad m \in \mathds{Z}+\lambda+\tfrac{1}{2}\ .
\ee
Here the penultimate entry in the ket indicates the transformation properties under the R-symmetry $\mathfrak{su}(2)_\mathrm{R}$, while the last entry describes the representation with respect to the outer automorphism. Thus these states transform in the representation $(\mathscr{C}^{j-\frac{1}{2}}_{\lambda+\frac{1}{2}},\mathbf{1})$ with respect to $\mathfrak{sl}(2,\mathds{R})\oplus \mathfrak{su}(2)_\mathrm{R}$, but are the highest weight states of a non-trivial representation of the outer automorphism.

We can now find the other vertex operators of the $\mathfrak{psu}(1,1|2)$ multiplet by applying the supercharges $S_0^{\alpha\beta-}$. First we consider the four states $S_0^{\alpha\beta-} |j,m,0, \uparrow \rangle$. Diagonalising the $\mathfrak{sl}(2,\mathds{R})\oplus \mathfrak{su}(2)_\mathrm{R}$ action leads to the linear combinations 
\begin{subequations}
\begin{align}
|j,m,\uparrow,0 \rangle&=S_0^{++-} |j,m-\tfrac{1}{2},0,\uparrow \rangle+S_0^{-+-} |j,m+\tfrac{1}{2},0,\uparrow \rangle\ , \\
|j,m,\uparrow,0 \rangle'&=(-j+m+1)S_0^{++-} |j,m-\tfrac{1}{2},0,\uparrow \rangle+(j+m-1)S_0^{-+-} |j,m+\tfrac{1}{2},0,\uparrow \rangle\ .
\end{align}
\end{subequations}
The states in the first line are the spin up component (with respect to $\mathfrak{su}(2)_\mathrm{R}$) of 
the representation $(\mathscr{C}^j_\lambda,\mathbf{2})$, 
whereas the states in the second line are the spin up component of $(\mathscr{C}^{j-1}_{\lambda},\mathbf{2})$. In the explicit Wakimoto representation, we thus have
\begin{subequations}
\begin{align}
|j,m ,\uparrow,0\rangle&=\big(j-\tfrac{3}{2}\big) \oint \mathrm{d}z\ \Big(\theta^{++} \gamma^{-j-m+1}+\theta^{-+}\gamma^{-j-m}\Big) \mathrm{e}^{(j-\frac{1}{2}) \sqrt{\frac{2}{k}}\Phi}(z) |0 \rangle\ , \label{eq:first descendant}\\
|j,m ,\uparrow,0\rangle'&=\big(j-\tfrac{1}{2}\big) \oint \mathrm{d}z\ \Big(\big(j-m-1\big)\theta^{++} \gamma^{-j-m+1}\nonumber\\
&\qquad\qquad\qquad\qquad-\big(j+m-1\big)\theta^{-+}\gamma^{-j-m}\Big) \mathrm{e}^{(j-\frac{1}{2}) \sqrt{\frac{2}{k}}\Phi}(z) |0 \rangle\ . \label{eq:second descendant}
\end{align}
\end{subequations}
We can similarly construct the vertex operators for the other descendants.

\subsection{The short representation} \label{subsec:short representation}

As an aside we should mention that the above multiplet shortens for $j=\tfrac{1}{2}$, which was one of the key insights in \cite{Eberhardt:2018ouy}.\footnote{For $j=\tfrac{3}{2}$, the multiplet also shortens, since \eqref{eq:first descendant} becomes null. However, the two representations are actually isomorphic since 
$(\mathscr{C}^{0}_{\lambda+\frac{1}{2}},\mathbf{1}) \cong (\mathscr{C}^{1}_{\lambda+\frac{1}{2}},\mathbf{1})$.} Indeed, \eqref{eq:second descendant} becomes null for $j=\tfrac{1}{2}$. In this special case, the $\mathfrak{psu}(1,1|2)$ representation is particularly simple and the complete list of vertex operators reads (we suppress the label $j=\tfrac{1}{2}$ in the following):
\begin{subequations}
\begin{align}
|m,0, \uparrow \rangle&=\oint \mathrm{d}z\ \gamma^{-m}|0\rangle\ , \\
|m ,\uparrow,0\rangle&=-\oint \mathrm{d}z\ \Big(\theta^{++} \gamma^{-m+\frac{1}{2}}+\theta^{-+}\gamma^{-m-\frac{1}{2}}\Big) |0 \rangle\ , \\
|m ,\downarrow,0\rangle&=-\oint \mathrm{d}z\ \Big(\theta^{+-} \gamma^{-m+\frac{1}{2}}+\theta^{--}\gamma^{-m-\frac{1}{2}}\Big) |0 \rangle\ , \\
|m ,0,\downarrow\rangle&=\oint \mathrm{d}z\ \Big(-\theta^{--}\theta^{-+} \gamma^{-m-1}+\theta^{-+}\theta^{+-}\gamma^{-m}\nonumber\\
&\qquad\qquad\qquad\qquad\qquad-\theta^{--}\theta^{++}\gamma^{-m}-\theta^{+-}\theta^{++} \gamma^{-m+1}\Big) |0 \rangle\ .
\end{align} 
\end{subequations}
We should stress that this shortening happens regardless of the value of $k$. While this multiplet is usually not part of the string theory spectrum as conjectured in \cite{Maldacena:2000hw}, at $k=1$ it is the only consistent multiplet \cite{Eberhardt:2018ouy}. This follows from the fact that any long multiplet contains a spin $\geq 1$ representation of R-symmetry $\mathfrak{su}(2)_k$, which is not allowed at level $k=1$. 

\subsection{Spectral flow} \label{subsec:spectral flow susy}
We will also need the behaviour of the Wakimoto representation under the spectral flow $\sigma$ of $\mathfrak{psu}(1,1|2)_k$ (which acts both on $\mathfrak{sl}(2,\mathds{R})_k$ and $\mathfrak{su}(2)_k$). The generating fields transform under spectral flow  as
\begin{subequations}
\begin{align}
\sigma^w(p^{\alpha\beta})(z)&=z^{\frac{1}{2}w(\beta-\alpha)} p^{\alpha\beta}(z)\ , \\
\sigma^w(\theta^{\alpha\beta})(z)&=z^{-\frac{1}{2}w(\beta-\alpha)} \theta^{\alpha\beta}(z)\ , \\
\sigma^w(\beta)(z)&=\beta(z)z^{-w}\ , \\
\sigma^w(\gamma)(z)&=\gamma(z)z^w\ , \label{wgam} \\
\sigma^w(\partial\Phi)(z)&=\partial \Phi(z)+\sqrt{\tfrac{k}{2}} \tfrac{w}{z}\ , \\
\sigma^w(\mathscr{K}^\pm)(z)&=\mathscr{K}^\pm(z)z^{\pm w}\ , \\
\sigma^w(\mathscr{K}^3)(z)&=\mathscr{K}^3(z)+\tfrac{kw}{2z}\ .
\end{align}
\end{subequations}

\section{The spacetime symmetry algebra} \label{sec:spacetime symmetry algebra}

In this Section we define the boundary symmetry generators (i.e.\ the DDF operators) for the background $\mathrm{AdS}_3 \times \mathrm{S}^3 \times \mathbb{T}^4$ following \cite{Giveon:1998ns,Ito:1998vd, Andreev:1999nt, Ashok:2009jw}. 
We will first show how to construct the operators that realise an extended spacetime $\mathcal{N}=4$ algebra. 

In the following it will be convenient to treat the torus excitations (that are independent of $k$) in the RNS formalism, while the $\mathrm{AdS}_3 \times \mathrm{S}^3$ part will be analysed in the hybrid formalism. This then also remains well-defined at $k=1$ (where only the $\mathrm{AdS}_3 \times \mathrm{S}^3$ part becomes ill-defined in the RNS formalism).

\subsection{Spacetime operators in the RNS-formalism} \label{subsec:spacetime operators in RNS}

We begin by constructing the spacetime operators for the torus directions in the RNS-formalism. Since these are independent of $k$, there is no need to invoke the hybrid formalism for them. In any case, we can always rewrite them in terms of the hybrid variables if we want to. 

\subsubsection{Free bosons} \label{subsubsec:free bosons}

The spacetime operators for the free bosons of the torus are given in the canonical $(-1)$ picture as
\be 
\partial \mathcal{X}^{(-1)\alpha}_m=\oint \mathrm{d}z\ \lambda^\alpha \mathrm{e}^{-\phi} \gamma^m\ , \qquad \partial \bar{\mathcal{X}}^{(-1)\alpha}_m=\oint \mathrm{d}z\ \bar{\lambda}^\alpha \mathrm{e}^{-\phi} \gamma^m\ .\label{eq:free boson picture zero}
\ee
One readily shows that these operators are BRST invariant under the BRST charge \eqref{eq:BRST charge}. Indeed, since the integrand has conformal weight $h=1$, it is invariant under the bosonic piece $Q_0$ \eqref{eq:BRST charge Q0}. On the other hand, both $Q_1$ and $Q_2$ have regular OPEs with the integrand.

For the following it will also be convenient to evaluate these operators in the $(0)$ picture, where they become
\begin{align}
\partial \mathcal{X}^{(0)\alpha}_m&=-2\oint \mathrm{d}z \ G_{-1/2}(\lambda^\alpha \gamma^m)(z) \\
 &=\oint \mathrm{d}z \ \Big( \partial X^\alpha \gamma^m-\frac{m}{k} \lambda^\alpha\big(2\psi^3 \gamma^m-\psi^+ \gamma^{m+1}-\psi^- \gamma^{m-1}\big)\Big)(z)\ ,
\end{align}
and similarly for the barred bosons. These generators satisfy then the commutation relations\footnote{This is most easily evaluated by taking one of the generators in the $(-1)$ picture and the other in the $(0)$ picture, and then changing the picture, but one can also work this out directly with both of them in the $(0)$ picture.}
\begin{align} 
[\partial \mathcal{X}^{(0)\alpha}_m,\partial \bar{\mathcal{X}}^{(0)\beta}_n]&=-n\epsilon^{\alpha\beta}\mathcal{I}^{(0)}_{m+n}\ ,
\end{align}
where we have defined
\be  \label{I0m}
\mathcal{I}^{(0)}_{m}\equiv\oint \mathrm{d}z\  (\gamma^{m-1} \partial \gamma)(z)\ .
\ee
For $m \ne 0$, the integrand is a total derivative and vanishes, and thus $\mathcal{I}_m^{(0)}=\mathcal{I}\, \delta_{m,0}$, where $\mathcal{I}$ is the `identity' operator that was already introduced in the bosonic analysis, see eq.~(\ref{Idef}). Thus we arrive at 
\be 
[\partial \mathcal{X}^{(0)\alpha}_m,\partial \bar{\mathcal{X}}^{(0)\beta}_m]= m \delta_{m+n}\epsilon^{\alpha\beta}\mathcal{I}\ .
\ee 
Of course, this relation now holds in any picture. Similarly, one checks that 
\be 
[\partial \mathcal{X}^{(0)\alpha}_m,\partial \mathcal{X}^{(0)\beta}_m]=0\ , \qquad [\partial \bar{\mathcal{X}}^{(0)\alpha}_m,\partial \bar{\mathcal{X}}^{(0)\beta}_m]=0\ .
\ee

\subsubsection{Free fermions} \label{subsubsec:free fermions}
Next we come to the fermionic operators. The free fermions are best constructed in the canonical $(-\tfrac{1}{2})$ picture, where the relevant vertex operators are given as\footnote{Since they describe spacetime fermions they come from the R-sector on the world-sheet; their structure can then be read off from \eqref{eq:bosonisation a} and \eqref{eq:bosonisation b}.}
\begin{multline} 
\Lambda^{(-\frac{1}{2})\alpha}_r=k^{-\frac{1}{4}}\oint \mathrm{d} z\ \Big(\mathrm{e}^{\frac{1}{2}H_1+\frac{\alpha}{2} H_2-\frac{\alpha}{2} H_3+\frac{1}{2} H_4-\frac{1}{2} H_5-\frac{\phi}{2}}\gamma^{r+\frac{1}{2}} \\
+\mathrm{e}^{-\frac{1}{2}H_1+\frac{\alpha}{2} H_2+\frac{\alpha}{2} H_3+\frac{1}{2} H_4-\frac{1}{2} H_5-\frac{\phi}{2}}\gamma^{r-\frac{1}{2}}\Big)\ , \label{eq:torus fermions -1/2 picture}
\end{multline}
and similarly for the complex conjugates
\begin{multline} 
\bar{\Lambda}^{(-\frac{1}{2})\alpha}_r=k^{-\frac{1}{4}} \oint \mathrm{d} z\ \Big(\mathrm{e}^{\frac{1}{2}H_1+\frac{\alpha}{2} H_2-\frac{\alpha}{2} H_3-\frac{1}{2} H_4+\frac{1}{2} H_5-\frac{\phi}{2}}\gamma^{r+\frac{1}{2}} \\
+\mathrm{e}^{-\frac{1}{2}H_1+\frac{\alpha}{2} H_2+\frac{\alpha}{2} H_3-\frac{1}{2} H_4+\frac{1}{2} H_5-\frac{\phi}{2}}\gamma^{r-\frac{1}{2}}\Big)\ .\label{eq:torus fermions -1/2 picture complex conjugate}
\end{multline}
Here $\alpha$ will be an R-symmetry index, and we have suppressed the cocycle factors that are necessary for locality.
These vertex operators (anti)commute again trivially with $Q_0$ and $Q_2$. To demonstrate invariance with respect to $Q_1$, one has to show that $G(z)$ has no $(z-w)^{-3/2}$ singularity with the integrand. There are two potential contributions to this singularity, one arising from the cubic fermion terms in \eqref{eq:worldsheet supercurrent}, and one from the contraction of the $\mathfrak{sl}(2,\mathds{R})$ part in the first line of \eqref{eq:worldsheet supercurrent} with the $\gamma$'s in the integrand of (\ref{eq:torus fermions -1/2 picture}) or (\ref{eq:torus fermions -1/2 picture complex conjugate}). As it turns out the two contributions cancel precisely, thus proving that these operators are indeed BRST invariant.  Furthermore, since the exponents of the $H_i$ involve an even number of $(-)$-signs, the operators also respect the GSO projection.

The anticommutators of these fermionic operators can be computed directly, and one finds 
\begin{align}
\{\Lambda^{(-\frac{1}{2})\alpha}_r,\Lambda^{(-\frac{1}{2})\beta}_s\}&=0\ , \\
\{\bar{\Lambda}^{(-\frac{1}{2})\alpha}_r,\bar{\Lambda}^{(-\frac{1}{2})\beta}_s\}&=0\ , \\
\{\Lambda^{(-\frac{1}{2})\alpha}_r,\bar{\Lambda}^{(-\frac{1}{2})\beta}_s\}&=\frac{\epsilon^{\alpha\beta}}{k}\oint \mathrm{d}z\ \mathrm{e}^{-\phi}\Big(\psi^+ \gamma^{r+s+1}-2\psi^3 \gamma^{r+s}+\psi^- \gamma^{r+s-1}\Big)(z)
\\
&=\epsilon^{\alpha\beta} \mathcal{I}_{r+s}^{(-1)}\ ,
\end{align}
where we have used picture changing in the last step.\footnote{Strictly speaking, since we have not kept track of the cocycle factors, our calculation only shows that the last anti-commutator is given by the right-hand-side up to a sign (which could in principle also depend on $\alpha=-\beta$). However, given that $\alpha$ and $\beta$ are spinor indices with respect to the outer $\mathfrak{su}(2)$ symmetry, the dependence must be proportional to 
$\epsilon^{\alpha\beta}$.} Thus, the fermions behave as free fields in spacetime.

\subsection{The spacetime operators in the hybrid formalism} \label{subsec:spacetime operators hybrid formalism}

Next we want to construct the spacetime ${\cal N}=4$ algebra whose supercharges will transform these boson and fermion fields into one another. Since the superconformal symmetry arises from the ${\rm AdS}_3 \times {\rm S}^3$ part of the background, we should now switch to the hybrid formalism. 

\subsubsection{The Virasoro algebra} \label{subsubec:Virasoro algebra}

Let us begin with the spacetime Virasoro algebra that was already (in the zero picture) given in Section~\ref{sec:DDF}, see eq.~(\ref{eq:spacetime Virasoro algebra bos})
\begin{align} 
\mathcal{L}_n^{(0)}= \oint \mathrm{d}z\ \Big(\big(1-n^2) J^3 \gamma^n+\tfrac{n(n-1)}{2} J^+\gamma^{n+1}+\tfrac{n(n+1)}{2} J^-\gamma^{n-1}\Big)(z)\ . \label{eq:spacetime Virasoro algebra}
\end{align}
Incidentally, this formula is the same in the NSR and the hybrid formalism --- the calculations of \cite{Giveon:1998ns} were done in the RNS formalism --- since the $\mathfrak{sl}(2,\mathds{R})_k$ currents $J^a$ are the same in both descriptions.\footnote{These generators describe spacetime symmetries, and hence should agree.  One can also check this explicitly by inserting (\ref{pdef}) and (\ref{thetadef}) into (\ref{eq:Wakimoto representation J}).} A direct computation similar to the bosonic case shows that
\be 
[\mathcal{L}_m^{(0)},\mathcal{L}_n^{(0)}]=(m-n)\mathcal{L}_{m+n}^{(0)}+\frac{k}{2}m(m^2-1)\, \mathcal{I}^{(0)}_{m+n}\ ,
\ee
where $\mathcal{I}^{(0)}_m=\mathcal{I}\, \delta_{m,0}$ is again given by (\ref{I0m}). One also checks that the spacetime free fields from the torus transform as primary fields of conformal weight 1 and $\frac{1}{2}$ (for the bosons and fermions, respectively) with respect to this Virasoro algebra; this is the analogue of \eqref{eq:spacetime primary}.

\subsubsection{The supercharges} \label{subsubec:supercharges hybrid}

Next we want to find the DDF operators for the supercharges. In the $(-\tfrac{1}{2})$ picture they are given as 
\begin{multline}
\mathcal{G}_r^{(-\frac{1}{2})\alpha\beta}=k^{\frac{1}{4}}\oint \mathrm{d}z\ \Big(\big(r-\tfrac{1}{2}\big) \mathrm{e}^{\frac{1}{2}H_1+\frac{\alpha}{2} H_2+\frac{\alpha}{2} H_3+\frac{\beta}{2} H_4+\frac{\beta}{2} H_5-\frac{\phi}{2}}\gamma^{r+\frac{1}{2}} \\
+\big(r+\tfrac{1}{2}\big) \mathrm{e}^{-\frac{1}{2}H_1+\frac{\alpha}{2} H_2-\frac{\alpha}{2} H_3+\frac{\beta}{2} H_4+\frac{\beta}{2} H_5-\frac{\phi}{2}}\gamma^{r-\frac{1}{2}}\Big)(z)\ . \label{eq:spacetime supercharges RNS}
\end{multline}
We have written this formula in terms of the RNS fields since then the expressions are simpler (and more symmetrical). Note that for the $\mathcal{G}^{(-\frac{1}{2})\alpha+}_{\mp \beta/2}$ generators, the integrand is equal to $p^{\beta\alpha}$, see (\ref{pdef}), and hence these generators also have a simple description in the hybrid formalism; the $\mathcal{G}^{\alpha-}_{\mp \beta/2}$ generators are more easily described in terms of the hybrid fields in the $(+\frac{1}{2})$ picture, see below. 

One can again check that these operators commute with the BRST charge and preserve the GSO-projection. Furthermore, they indeed transform the bosonic and fermionic spacetime operators into one another, i.e.\
\begin{align}
\{\mathcal{G}_r^{(-\frac{1}{2})\alpha\beta},\Lambda^{(-\frac{1}{2})\gamma}_s\}&=\epsilon^{\alpha\gamma}\partial \mathcal{X}^{(-1)\beta}_{r+s}\ , \\
[\mathcal{G}_r^{(-\frac{1}{2})\alpha\beta},\partial \mathcal{X}^{(0)\gamma}_m]&=m \epsilon^{\beta\gamma} \Lambda_{m+r}^{(-\frac{1}{2})\alpha}\ ,
\end{align}
and similarly for the barred free fields.
\medskip

In order to confirm that the supercharges generate the ${\cal N}=4$ superconformal algebra, it is convenient to take the supercharges $\mathcal{G}^{\alpha+}_r$ in the $(-\tfrac{1}{2})$ picture, and the supercharges $\mathcal{G}^{\alpha-}_r$ in the 
$(+\tfrac{1}{2})$ picture. Applying picture changing on $\mathcal{G}^{(-\frac{1}{2})\alpha-}_{\pm \frac{1}{2}}$ gives rise to \eqref{eq:Wakimoto representation S-} with two extra terms,\footnote{Thus  the Wakimoto representation \eqref{eq:Wakimoto representation S-} essentially implements picture changing.}
\be \label{Gp12}
\mathcal{G}_{\pm \frac{1}{2}}^{(+\frac{1}{2})\alpha-}=\pm \tilde{S}_0^{\mp\alpha-}\equiv\pm S_0^{\mp\alpha-}\pm\oint \mathrm{d}z\ \Big(p^{\mp\alpha} \mathrm{e}^\rho(\partial \bar{X}^-\Psi^+-\partial X^-\bar{\Psi}^+)+\frac{1}{2}p^{\mp \alpha} b \mathrm{e}^\rho\Big)\ ,
\ee
where we have now written the generators in terms of the hybrid fields. We should mention that the two extra terms will not modify the $\mathfrak{psu}(1,1|2)_k$ algebra itself, but couple it to the free bosons and fermions of $\mathbb{T}^4$.

The expression for general mode number $r$ can be obtained by taking the commutator with the Virasoro generators, and one finds 
\begin{multline}
\mathcal{G}_r^{(+\frac{1}{2})\alpha-}=\oint \mathrm{d}z\ \Big(\big(r+\tfrac{1}{2}\big) \tilde{S}^{-\alpha-} \gamma^{r-\frac{1}{2}}+\big(r-\tfrac{1}{2}\big) \tilde{S}^{+\alpha-}\gamma^{r+\frac{1}{2}} \\
+\big(r^2-\tfrac{1}{4}\big) \big(2J^3\theta^{-\alpha} \gamma^{r-\frac{1}{2}}+2J^3\theta^{+\alpha}\gamma^{r+\frac{1}{2}} -J^+ \theta^{-\alpha} \gamma^{r+\frac{1}{2}} \\
-J^- \theta^{-\alpha} \gamma^{r-\frac{3}{2}}-J^+ \theta^{+\alpha} \gamma^{r+\frac{3}{2}} -J^- \theta^{+\alpha} \gamma^{r-\frac{1}{2}} \big) \Big)\ ,
\end{multline}
i.e.\ there are further correction terms, see the contributions from the second and third line. Incidentally, this expression can also be obtained directly by applying picture changing to the expressions for the supercharges in the $(-\frac{1}{2})$ picture. 

\subsubsection{The spacetime $\mathfrak{su}(2)$-currents} \label{subsubsec:spacetime su2}

With these expressions at hand we can now calculate the anti-commutators of the supercharges, and thereby read off the form of the 
spacetime affine $\mathfrak{su}(2)$ algebra generators,
\begin{multline}\label{Kgen}
\mathcal{K}^{(0)a}_m=\oint \mathrm{d}z\ \Big(K^a \gamma^m-\tfrac{m}{2}\tensor{(\sigma^a)}{_{\alpha\beta}}\big(S^{+\alpha+} \theta^{+\beta} \gamma^{m+1}+S^{-\alpha+} \theta^{-\beta} \gamma^{m-1}\\
+S^{+\alpha+}\theta^{-\beta} \gamma^m+S^{-\alpha+}\theta^{+\beta} \gamma^m\big) \Big)\ .
\end{multline}
This agrees with what one obtains from \cite{Giveon:1998ns} upon rewriting the RNS fields in terms of the hybrid fields.

\subsection{The complete spacetime algebra} \label{subsec:complete spacetime algebra}

It remains to check that the generators (\ref{eq:spacetime Virasoro algebra}), (\ref{eq:spacetime supercharges RNS}), (\ref{Gp12}) and (\ref{Kgen}) satisfy the (anti-)commutation relations of the small $\mathcal{N}=4$ algebra, 
\begin{subequations}
\begin{align}
[\mathcal{L}_m,\mathcal{L}_n]&=\tfrac{k}{2}\, \mathcal{I}\,m(m^2-1) \delta_{m+n,0}+(m-n)\mathcal{L}_{m+n}\ , \label{eq:spacetime algebra a}\\
[\mathcal{L}_m,\mathcal{G}^{\alpha\beta}_r]&=\big(\tfrac{1}{2}m-r\big)\mathcal{G}^{\alpha\beta}_{m+r}\ , \label{eq:spacetime algebra b}\\
[\mathcal{L}_m,\mathcal{K}^a_n]&=-n \mathcal{K}^a_{m+n}\ , \label{eq:spacetime algebra c}\\
[\mathcal{K}^3_m,\mathcal{K}^3_n]&=\tfrac{k}{2}\, \mathcal{I}\,m \delta_{m+n,0}\ , \label{eq:spacetime algebra d}\\
[\mathcal{K}^3_m,\mathcal{K}^\pm_n]&=\pm\mathcal{K}^\pm_{m+n}\ , \label{eq:spacetime algebra e}\\
[\mathcal{K}^+_m,\mathcal{K}^-_n]&=k\, \mathcal{I}\,m \delta_{m+n,0}+2\mathcal{K}^3_{m+n}\ , \label{eq:spacetime algebra f}\\
[\mathcal{K}^a_m,\mathcal{G}^{\alpha\beta}_r]&=\tfrac{1}{2}\tensor{(\sigma^a)}{^\alpha_\gamma}\mathcal{G}^{\gamma\beta}_{r+m}\ , \label{eq:spacetime algebra g}\\
\{\mathcal{G}^{\alpha\beta}_r,\mathcal{G}^{\gamma\delta}_{s}\}&=k\big(r^2-\tfrac{1}{4}\big) \epsilon^{\alpha\gamma}\epsilon^{\beta\delta}\mathcal{I}\,\delta_{r+s,0}+\epsilon^{\alpha\gamma}\epsilon^{\beta\delta}\mathcal{L}_{r+s}+(r-s)\epsilon^{\beta\delta} (\sigma_a)^{\alpha\gamma} \mathcal{K}^a_{r+s}\ , \label{eq:spacetime algebra h}
\end{align}
\end{subequations}
and this turns out to be the case. Furthermore, the free boson and fermion fields extend this algebra to the so-called extended small ${\cal N}=4$ algebra, 
\begin{subequations}
\begin{align}
\{\mathcal{G}^{\alpha\beta}_r,\Lambda^\gamma_s\}&=\epsilon^{\alpha\gamma}(\partial \mathcal{X})^\beta_{r+s}\ , \label{eq:spacetime algebra i}\\
[\mathcal{G}^{\alpha\beta}_r,\partial\mathcal{X}^\gamma_m]&=m\epsilon^{\beta\gamma}\Lambda^\alpha_{r+m}\ , \label{eq:spacetime algebra j}\\
\{\mathcal{G}^{\alpha\beta}_r,\bar{\Lambda}^\gamma_s\}&=\epsilon^{\alpha\gamma}(\partial \bar{\mathcal{X}})^\beta_{r+s}\ , \label{eq:spacetime algebra k}\\
[\mathcal{G}^{\alpha\beta}_r,\partial\bar{\mathcal{X}}^\gamma_m]&=m\epsilon^{\beta\gamma}\bar{\Lambda}^\alpha_{r+m}\ , \label{eq:spacetime algebra l}\\
[\partial \mathcal{X}^\alpha_m,\partial \bar{\mathcal{X}}^\beta_n]&= m \epsilon^{\alpha\beta}\,\mathcal{I}\, \delta_{m+n,0}\ , \label{eq:spacetime algebra m}\\
\{\Lambda^\alpha_r,\bar{\Lambda}^\beta_s\}&= \epsilon^{\alpha\beta}\, \mathcal{I}\, \delta_{r+s,0}\ . \label{eq:spacetime algebra n}
\end{align}
\end{subequations}
We have checked these relations in specific pictures, but they remain then also true in general (and we have therefore not written the picture numbers explicitly). We should also mention that the identity operator $\mathcal{I}$ commutes with the ${\cal N}=4$ generators and the free fields, and that the free fields transform as Virasoro primaries.

\subsection{The action of the spacetime algebra on physical states} \label{sec:spacetime action}

As in the bosonic case discussed in Section~\ref{subsec:locality and moding} it remains to understand the values the various mode numbers can take. The argument that was given there continues to hold essentially unmodified --- the $\gamma$ field has the property that $\gamma^m$ is single-valued for any $m\in \mathds{R}$, see eq.~(\ref{gammam}). Given the form of the spectral flow on $\gamma$, see eq.~(\ref{wgam}), as well as the form of the various DDF operators, it follows that the mode numbers of bosonic DDF operators may be taken to lie in\footnote{For $w=0$ the modes can a priori take any real value.}
\be\label{wbos susy}
\hbox{$w$-th spectrally flowed sector:} \qquad 
n \ \in \tfrac{1}{w} \mathds{Z} \ ,
\ee
while the condition for the fermionic generators is instead 
\be\label{wfer susy}
\hbox{$w$-th spectrally flowed sector:} \qquad 
 r \in \tfrac{1}{w} \mathds{Z}+\tfrac{1}{2} \ . 
\ee
This is again reminiscent of the fractionally moded algebra in the symmetric orbifold \cite{Lunin:2001pw}, and indeed the untwisting that was done in the bosonic case, see Section~\ref{subsec:untwisting}, can be similarly performed. 
As in the bosonic case, the DDF operators map then different continuous representations $\mathscr{C}^j_\lambda$ into one another, see eq.~(\ref{2.31}). While both $\mathscr{C}^j_{\lambda} $ and $\mathscr{C}^j_{\lambda-n} $ are part of the world-sheet spectrum, there is actually a non-trivial constraint in that in the `diagonal modular invariant' we are considering, only the 
combinations $\mathscr{C}^j_\lambda \otimes \mathscr{C}^j_\lambda$ appear in the Hilbert space, i.e.\ the left- and right-moving $\lambda$ always agree. This means that, in order to map physical states to physical states, we need to combine left- and right-moving DDF operators such that the total left- and right-moving mode numbers differ by an integer (for bosonic DDF operators). This condition reflects precisely the orbifold invariance condition of the spacetime CFT  \cite{Giribet:2018ada,Gaberdiel:2018rqv,Eberhardt:2018ouy}.

\section{The symmetric product orbifold} \label{sec:symmetric product orbifold}

As we have seen above, the spectrally flowed continuous world-sheet representations give rise to the different single-cycle twisted sectors of a symmetric orbifold. In the previous section we have identified some of the generators of the corresponding seed theory; in particular, we have shown that the seed theory contains the extended small ${\cal N}=4$ superconformal algebra with $c=6k$, see eq.~(\ref{eq:spacetime algebra a}). In this section we show that the dual spacetime CFT is the symmetric orbifold  
\be 
\text{Sym}^N\Bigl(
\Bigl[\text{small }\mathcal{N}=4\text{ Liouville with }c=6(k-1)\Bigr] \oplus\ \mathbb{T}^4 \Bigr)\  . \label{eq:seed theory}
\ee
In particular, for $k=1$, the Liouville part vanishes and we recover the symmetric orbifold of $\mathbb{T}^4$ \cite{Eberhardt:2018ouy}, see Section~\ref{subsec:k1 case} below. 

\subsection{The \texorpdfstring{$\mathbb{T}^4$}{T4} algebra} \label{subsec:T4 algebra}

The first step of our argument consists of separating out the torus degrees of freedom from the rest (which is the $\mathcal{N}=4$ analogue of \eqref{eq:bosonic decoupling Virasoro}). In order to do so, we note that we can construct small $\mathcal{N}=4$ generators out of the free fields as 
\begin{subequations}
\begin{align}
(\mathcal{K}_{\mathbb{T}^4}^a)_m&=\tfrac{1}{2}(\sigma^a)_{\alpha\beta} (\Lambda^\alpha \bar{\Lambda}^\beta)_m\ , \\
(\mathcal{G}_{\mathbb{T}^4}^{\alpha\beta})_r&= (\Lambda^\alpha \partial \bar{\mathcal{X}}^\beta)_r-(\bar{\Lambda}^\alpha \partial \mathcal{X}^\beta)_r\ , \\
(\mathcal{L}_{\mathbb{T}^4})_m&= \epsilon_{\alpha\beta} (\partial \mathcal{X}^\alpha \partial \bar{\mathcal{X}}^\beta)_m+\tfrac{1}{2} \epsilon_{\alpha\beta}\big((\partial \Lambda^\alpha \bar{\Lambda}^\beta)_m-(\Lambda^\alpha \partial\bar{\Lambda}^\beta)_m\big)\ .
\end{align}
\end{subequations}
These generators satisfy the small $\mathcal{N}=4$ algebra with $c=6$.
We can then decouple the $\mathbb{T}^4$ part from the small $\mathcal{N}=4$ algebra by considering the differences $\mathcal{L}_m-(\mathcal{L}_{\mathbb{T}^4})_m$ and similarly for $\mathcal{G}_r^{\alpha\beta}$ and $\mathcal{K}^a_m$. These differences satisfy again the small $\mathcal{N}=4$ algebra, but commute with the torus modes. This shows that the chiral algebra of the seed theory of the symmetric product orbifold is
\be 
\Bigl[ \text{small }\mathcal{N}=4\text{ with }c=6(k-1)\Bigr]  \oplus\ \text{4 free bosons and 4 free bosons}\ . \label{eq:spacetime algebra}
\ee

\subsection{\texorpdfstring{$\mathcal{N}=4$}{N=4} Liouville theory} \label{subsec:N4 Liouville theory}

Next we want to show that the first term should be thought of as $\mathcal{N}=4$ Liouville theory. Since $\mathcal{N}=4$ Liouville theory is not very well known, we briefly review its main features below. 

$\mathcal{N}=4$ Liouville theory can be constructed starting from an $\mathcal{N}=1$ supersymmetric WZW-model based on $\mathrm{SU}(2)\times \mathrm{U}(1)$  \cite{Sevrin:1988ew, Ketov:1996es, Seiberg:1999xz, Eguchi:2016cyh}, together with some background charge for the $\mathrm{U}(1)$ factor so that the total central charge is $c=6\kappa$. (For the application we have in mind, we will later identify $\kappa= k-1$.)  We denote the generating fields by\footnote{These fields should not be confused with the world-sheet fields we were describing earlier: from now on we shall only talk about the fields of the (seed theory) of the spacetime CFT. We use the same symbols as before since this is the usual convention for ${\cal N}=4$ Liouville theory.}
\be \label{Liouvillefields}
\psi^{\alpha\beta}\ , \quad \partial\phi \quad\text{and}\quad J^a\ .
\ee
Here, $\partial \phi$ is the bosonic generator of the $\mathrm{U}(1)$ factor. The indices $\alpha$ and $\beta$ are spinor indices while $a$ is an adjoint index of $\mathfrak{su}(2)$, and the currents $J^a$ generate the affine $\mathfrak{su}(2)$ algebra at level $\kappa-1$. Our conventions for their OPEs can be found in Appendix~\ref{subapp:N4 Liouville conventions}. The chiral algebra has actually large ${\cal N}=4$ superconformal symmetry, but it also contains a small ${\cal N}=4$ algebra \cite{Sevrin:1988ew, Eguchi:2016cyh}, whose generators take the form 
\begin{subequations}
\begin{align}
T&=\frac{1}{\kappa+1}\Big(J^3J^3+\tfrac{1}{2}\big(J^+J^-+J^-J^+\big)\Big)+\frac{1}{2}\epsilon_{\alpha\gamma}\epsilon_{\beta\delta} \partial \psi^{\alpha\beta}\psi^{\gamma\delta}\nonumber\\
&\qquad\qquad\qquad\qquad\qquad\qquad\qquad\qquad+\frac{1}{2}\partial \phi \partial \phi+\frac{i \kappa}{\sqrt{2(\kappa+1)}} \partial^2\phi\ , \\
G^{\alpha\beta}&=\frac{1}{\sqrt{2}} (\partial\phi \psi^{\alpha\beta})+\frac{i}{\sqrt{\kappa+1}} \Big(-\tensor{(\sigma_a)}{^\alpha_\gamma} \big(J^a+\tfrac{1}{3}J^{(\text{f},+)a}\big)\psi^{\gamma\beta}\nonumber\\
&\qquad\qquad\qquad\qquad\qquad\qquad\qquad\qquad+\tfrac{1}{3} \tensor{(\sigma_a)}{^\beta_\gamma} J^{(\text{f},-)a}\psi^{\alpha\gamma}+\kappa \partial \psi^{\alpha\beta}\Big)\ , \\
K^a&=J^a+J^{(\text{f},+)a}\ ,
\end{align}
\end{subequations}
where the fermionic currents are
\be 
J^{(\text{f},+)a}=\frac{1}{4} \tensor{(\sigma^a)}{_{\alpha\gamma}}\epsilon_{\beta\delta} (\psi^{\alpha\beta}\psi^{\gamma\delta})\quad\text{and}\quad 
J^{(\text{f},-)a}=\frac{1}{4} \epsilon_{\alpha\gamma}\tensor{(\sigma^a)}{_{\beta\delta}} (\psi^{\alpha\beta}\psi^{\gamma\delta})\ .
\ee
The vertex operators of this theory can be described by 
\be 
\mathrm{e}^{\frac{1}{\sqrt{2(\kappa+1)}}\left(2 p-i\kappa\right)\phi} \mathscr{V}_\ell\ , \label{eq:vertex operators}
\ee
where $p \in \mathds{R}$, and $\mathscr{V}_\ell$ denotes the $\mathfrak{su}(2)_{\kappa-1}$ primary of spin $\ell$.  It has conformal weight
\be 
h=\frac{\big(\ell+\frac{1}{2}\big)^2+p^2}{\kappa+1}+\frac{\kappa-1}{4}\ .
\ee
In particular, the lowest conformal dimension of spin $\ell$ is therefore 
\be 
\Delta_\ell^{\text{NS}}=\frac{\big(\ell+\frac{1}{2}\big)^2}{\kappa+1}+\frac{\kappa-1}{4}\ , \label{eq:Liouville gap NS sector}
\ee
above which there is a continuum of conformal weights. Here the $\mathfrak{su}(2)$ spin $\ell$ takes values in $\{0,\tfrac{1}{2},\dots,\tfrac{\kappa-1}{2}\}$, in agreement with the unitarity bound for the small ${\cal N}=4$ superconformal algebra \cite{Eguchi:1987sm}. 

The above discussion applies to the NS-sector. In the R-sector  there is an additional contribution from the ground state energy of the four fermions, whose zero modes generate the $\mathfrak{su}(2)$ representation $\mathbf{2} \oplus \, 2\cdot \mathbf{1}$. As a consequence, the value of the $\mathfrak{su}(2)$ spin $\ell$ is shifted by one half, and the gap in the R-sector is
\be 
\Delta_\ell^{\text{R}}=\frac{\ell^2+p^2}{\kappa+1}+\frac{\kappa-1}{4}+\frac{1}{4}\ . \label{eq:Liouville gap R sector}
\ee
In the R-sector, the $\mathfrak{su}(2)$ spin $\ell$ takes values in $\{\tfrac{1}{2},1,\dots,\tfrac{\kappa}{2}\}$.

$\mathcal{N}=4$ Liouville theory has then the same spectrum as the above free boson theory, except that the operators associated to $p$ and $-p$ should be identified. (This will reflect the fact that for the continuous representations of $\mathfrak{sl}(2,\mathds{R})$, the two cases $j=\frac{1}{2} \pm i p$ define the same representation.) As a consequence, also the correlation functions of Liouville theory are more complicated than those calculated from the free boson theory.

\subsection{Identifying Liouville theory on the world-sheet}\label{sec:Identifys}

In the bosonic case, Liouville theory is believed to be uniquely characterised by having Virasoro symmetry, together with the full spectrum of Liouville fields \cite{Ribault:2014hia, Collier:2017shs}. It is tempting to speculate that a similar statement should be true for $\mathcal{N}=4$ Liouville theory. Since we have already shown that the first factor has small ${\cal N}=4$ superconformal symmetry, it only remains to show that the world-sheet theory gives rise to the full spectrum of Liouville theory. Since the single-particle perturbative part of the spacetime theory only has a NS-sector, we need to show that the spacetime spectrum exhibits the gaps (\ref{eq:Liouville gap NS sector}), together with a continuum above. Similarly, we need to show that the same is true for the twisted sectors of the symmetric orbifold (where for even twist $w$ we also need the R-sector ground state energy \eqref{eq:Liouville gap R sector}, see \cite{Gaberdiel:2018rqv}.)
This mirrors then precisely the analysis of Section~\ref{sec:Identify} for the bosonic case.

In order to establish this, we first note that the restriction of the $\mathfrak{su}(2)$ spins to $\ell \in \{0,\tfrac{1}{2},\dots,\tfrac{k-2}{2}\}$ is correctly implemented in the world-sheet theory, since the bosonic $\mathfrak{su}(2)$ algebra (which we denoted by $\mathscr{K}^a$ above) is at level $k-2$. To determine the gap predicted by the world-sheet theory, we simply have to solve the mass-shell condition on the worldsheet for a continuous representation. In the $w$ spectrally flowed NS-sector, it takes the form \cite{Gaberdiel:2018rqv}
\be 
\frac{\frac{1}{4}+p^2}{k}-hw+\frac{kw^2}{4}+\frac{\ell(\ell+1)}{k}=\frac{1}{2}\ ,
\ee
where the first term comes from the Casimir of the $\mathfrak{sl}(2,\mathds{R})_{k+2}$ representation, the next two terms arise from the spectral flow and the last term is the $\mathfrak{su}(2)_{k-2}$ ground state energy. (We are considering here the ground state with $N=0$.) Finally, the right-hand side is the appropriate normal ordering constant in the NS-sector. Here, $h$ denotes the spectrally flowed $J^3_0$ eigenvalue, which corresponds to the conformal weight in the dual CFT. 
From this, we solve
\be 
h=\frac{\big(\ell+\frac{1}{2}\big)^2+p^2}{wk}+\frac{kw^2-2}{4w}=\frac{6k}{24w}(w^2-1)+\frac{\Delta_\ell^\text{NS}}{w}+\frac{p^2}{wk}\ ,
\ee
which matches precisely with \eqref{eq:Liouville gap NS sector}, provided that $\kappa=k-1$. 
The ground state with $N=0$ is allowed to exist only for an odd unit of spectral flow, because of the GSO projection.\footnote{Here we have only spectrally flow in the $\mathfrak{sl}(2,\mathds{R})$ sector; then the GSO projection depends on the cardinality of the sepctral flow, see e.g.\ \cite{Ferreira:2017pgt}.}

For even spectral flow, we have to apply one fermion, which we take to be $\chi^+_{-1/2}$, i.e.\ the positively charged $\mathfrak{su}(2)$ fermion. The mass-shell condition then reads
\be
\frac{\frac{1}{4}+p^2}{k}-hw+\frac{k w^2}{4}+\frac{\ell(\ell-1)}{k}+\frac{1}{2}=\frac{1}{2}\ ,
\ee
where $\ell$ is the actual $\mathfrak{su}(2)$ spin of the state. (Since we have applied $\chi^+_{-1/2}$, it differs by one unit from the spin of the ground state, which is therefore $\ell_0 = \ell-1$.)  Solving the mass-shell condition yields now
\be \label{evenspec}
h=\frac{\ell(\ell-1)+\frac{1}{4}+p^2}{kw}+\frac{kw}{4}=\frac{6k}{24w}(w^2-1)+\frac{\Delta_{\ell-\frac{1}{2}}^\text{R}}{w}+\frac{p^2}{wk}+\frac{1}{4w}\ .
\ee
This matches with the R-sector ground state energy of Liouville theory, using that the symmetric orbifold in even twist sectors behaves effectively as in the R-sector, see \cite{Lunin:2001pw, David:2002wn}. Note that the  additional contribution $+\tfrac{1}{4w}$ in (\ref{evenspec}) comes from the fact that also the additional $\mathbb{T}^4$ in \eqref{eq:seed theory} is now in the R-sector for which the ground state energy is $\tfrac{1}{4}$. Furthermore, the $\mathfrak{su}(2)$ spin $\ell$ is shifted by $\tfrac{1}{2}$ with respect to pure Liouville theory, because of the additional zero modes of the torus theory $\mathbb{T}^4$. Thus, the representation content matches exactly.

\subsection{Spectrum generating algebra}

So far we have only shown that the $w$-spectrally flowed continuous representations give rise to a spacetime spectrum on which the $w$-cycle twisted sector operators of the generators in (\ref{eq:seed theory}) act. Now we want to show that these twisted sector operators generate in fact the entire spectrum. 

We shall first consider the case $k\geq 2$; the case $k=1$ will be discussed separately in the following section. 
For $k\geq 2$, the argument works essentially as in flat space. For $k\geq 2$, we have 8 bosonic and fermionic DDF operators as follows. For the bosonic operators, $4$ of them come from ${\cal N}=4$ Liouville theory (namely from the $J^a$ and $\partial \phi$ in (\ref{Liouvillefields})), while the other $4$ are the $4$ torus modes. For the fermions, we have $4$ fermionic generators from ${\cal N}=4$ Liouville theory (namely the $\psi^{\alpha\beta}$ in (\ref{Liouvillefields})), while the other $4$ generators are the $4$ torus fermions. 

These DDF operators can now be compared to the world-sheet description. The matching of the fermions is straightforward since they define free fields (and hence do not contain any null-vectors). As regards the bosons, we have before imposing the physical state conditions, 
$10$ bosonic generators on the world-sheet: $3$ from $\mathfrak{sl}(2)_{k+2}$, $3$ from $\mathfrak{su}(2)_{k-2}$, and $4$ from the torus. $\mathfrak{sl}(2,\mathds{R})_{k+2}$ does not have any null-vectors (for $k\geq 2$), and hence the physical state condition removes two of the bosonic generators, leaving essentially one boson behind (that we may identify with $\partial \phi$ in (\ref{Liouvillefields})). The $\mathfrak{su}(2)_{k-2}$ generators of the world-sheet can be directly identified with the $\mathfrak{su}(2)_{\kappa-1}$  generators of (\ref{Liouvillefields}) --- in particular, their characters agree precisely, including null-vectors --- while the remaining $4$ bosons are torus bosons in both descriptions. The fact that our DDF operators generate the entire spectrum then follows by the usual character argument. This is to say, we can easily calculate the character of the physical spectrum from the world-sheet, and it manifestly agrees with the corresponding character of the DDF operators. This works separately for each $w$, and for each ground-state representation. Thus the DDF operators we have constructed generate the full spacetime spectrum.

\subsection{The case of \texorpdfstring{$k=1$}{k=1}} \label{subsec:k1 case}

The case $k=1$ is very special. In particular, the Liouville part of the seed theory \eqref{eq:spacetime algebra}
now has $c=0$, and we would expect that it disappears entirely from the spectrum. In fact, this is precisely in agreement with what was shown in \cite{Eberhardt:2018ouy}, where we determined the world-sheet characters at $k=1$, and demonstrated that they are generated by $4$ free bosons and fermions. Thus we should only expect to have $4+4$ DDF operators, and these are precisely the ones associated to $\mathbb{T}^4$ (that will always exist). Furthermore, at $k=1$ the continuous representations account for the complete worldsheet theory, since there are no discrete representations on the worldsheet in this case \cite{Eberhardt:2018ouy}. 

We should emphasise that relative to \cite{Eberhardt:2018ouy}, where `only' the spectrum was matched, we have now established that the algebraic structure of the spacetime theory is indeed that of the symmetric orbifold of $\mathbb{T}^4$: we have shown that the spacetime CFT contains the spectrum generating operators of the symmetric orbifold with the correct commutation relations. This essentially amounts to proving that the spacetime theory is indeed the symmetric orbifold of $\mathbb{T}^4$.

\section{Discussion} \label{sec:discussion}
In this paper, we have considered string theory on $\mathrm{AdS}_3$ with pure NS-NS flux. By considering a complete set of DDF operators, we have shown that the spacetime theory is given by a symmetric orbifold of Liouville theory together with the internal CFT. We have established this for bosonic string theory on $\mathrm{AdS}_3 \times X$, as well as for superstrings on $\mathrm{AdS}_3 \times \mathrm{S}^3 \times \mathbb{T}^4$; in the latter case, the dual CFT is the symmetric orbifold of the product of $\mathcal{N}=4$ Liouville theory with the $\mathbb{T}^4$ theory. We have moreover seen that the $k=1$ limit considered in \cite{Eberhardt:2018ouy} comes about naturally, since in this case the $\mathcal{N}=4$ Liouville part (together with its long string continuum) disappears.

This gives a fairly complete picture of holography on $\mathrm{AdS}_3$ with pure NS-NS flux. The background is indeed `singular', but this does not hinder the existence of a well-defined dual CFT. In the general case, the proposed dual CFTs contain also a continuum of states, and in particular the vacuum is non-normalisable. As we have seen, the entire spacetime spectrum is accounted for by the continuous representations on the world-sheet. We have argued in Section~\ref{subsec:discretereps} that the discrete representations on the world-sheet give rise to non-normalisable operators in the dual CFT that are not directly part of the CFT spectrum, see also \cite{Maldacena:2001km}.

While our discussion in the bosonic case was general, we focused on the specific example of $\mathrm{AdS}_3 \times \mathrm{S}^3 \times \mathbb{T}^4$ in the supersymmetric case. This is because the fermions couple the $\mathrm{AdS}_3$ factor to the rest of the background, and one cannot easily treat the general case uniformly. For instance, in the case of K3, the spacetime theory has also small $\mathcal{N}=4$ supersymmetry and at least at the orbifold point of $\mathrm{K3}$, one easily sees that the general answer for the dual CFT will be
\be 
\text{Sym}^N\left(\Bigl[\mathcal{N}=4\text{ Liouville with }c=6(k-1)\Bigr] \oplus\ \mathrm{K3}\right)\ .
\ee
In particular, for $k=1$, one simply recovers the symmetric orbifold of K3 \cite{Schmidthesis}.
Another interesting background is given by $\mathrm{AdS}_3 \times \mathrm{S}^3 \times \mathrm{S}^3 \times \mathrm{S}^1$ \cite{Elitzur:1998mm, Gukov:2004ym, Eberhardt:2017pty, Eberhardt:2017fsi}. We will show in \cite{LorenzMatthias} that the spacetime CFT turns out to be
\be 
\text{Sym}^N\left(\text{large }\mathcal{N}=4\text{ Liouville with }c=\frac{6k^+k^-}{k^++k^-}\right)\ ,
\ee
which collapses to the symmetric orbifold of $\mathrm{S}^3 \times \mathrm{S}^1$ if one of the fluxes through the spheres attains its minimal value $k^+=1$ or $k^-=1$. Similarly, the dual CFTs of the orbifold backgrounds \cite{Kutasov:1998zh, Datta:2017ert, Eberhardt:2017uup, Yamaguchi:1999gb, Eberhardt:2018sce} should simply be given by the orbifolds of the respective (extended) Liouville theories.

It is interesting to note that the seed theories of the dual CFTs we have given are essentially the Drinfel'd Sokolov (quantum Hamiltonian) reductions of the respective worldsheet theories, in close analogy to the higher spin setting \cite{Campoleoni:2010zq}. Indeed, it is well-known that the quantum Hamiltonian reduction of $\mathfrak{sl}(2,\mathds{R})_k$ yields Liouville theory with central charge \cite{Bouwknegt:1992wg}
\be 
c=13-6(-k+2)-\frac{6}{-k+2}=1+\frac{6(k-1)^2}{k-2}\ ,
\ee
which differs by 24 from \eqref{eq:bosonic Liouville central charge}. This is related to the fact that in bosonic string theory, the ghosts contribute central charge $c=-26$, whereas in the quantum Hamiltonian reduction, they only contribute $c=-2$. Thus, while our construction is certainly related to quantum Hamiltonian reduction, it is not exactly clear what the precise relation should be.

\section*{Acknowledgements} We thank Rajesh Gopakumar for many useful discussions. We also thank Andrea Dei, Sylvain Ribault and Thilo Schmid for useful conversations. LE is supported by the Swiss National Science Foundation, and the work of the group is more generally supported by the NCCR SwissMAP which is also funded by the Swiss National Science Foundation.

\appendix

\section{Higher spin fields in spacetime} \label{app:higher spins}

In this Appendix we explain how to construct DDF operators associated to the higher spin generators on the world-sheet. Since this construction does not seem to have been discussed in the literature before, we will be fairly explicit. 

\subsection{Internal Virasoro algebra}

Let us begin by constructing the DDF operators associated to the (internal) Virasoro algebra arising from $X$, whose Virasoro tensor we denote by  $T^\mathrm{m}(z)$. We define the corresponding spacetime Virasoro generators via 
\be 
\mathcal{L}_m^\text{m} \equiv \oint \mathrm{d}z\ \big((\partial \gamma)^{-1} \gamma^{m+1} T^\text{m}\big)(z)+\frac{c^\text{m}}{12} \oint \mathrm{d}z\ \gamma^{m+1} \left(\frac{3}{2}(\partial^2\gamma)^2 (\partial \gamma)^{-3}-\partial^3 \gamma(\partial \gamma)^{-2}\right)\ .\label{eq:internal Virasoro spacetime}
\ee
This definition is a bit formal, but as we shall see it makes sense. In particular, since $\partial \gamma$ is a primary field on the worldsheet, the first term in the definition is a quasi-primary field on the worldsheet. The second term corrects for the fact that $T^\text{m}$ is only quasi-primary and makes the expression primary. It is essentially the Schwarzian derivative of the transformation $z \mapsto \gamma(z)$. There are no normal-ordering ambiguities in the definition \eqref{eq:internal Virasoro spacetime}, since $\gamma(z)$ has regular OPE with itself (as well as with all its derivatives).

We can calculate the algebra of modes via 
\begin{align}
[\mathcal{L}^\text{m}_m,\mathcal{L}^\text{m}_n]&=\oint_0 \mathrm{d}w\ \oint_w \mathrm{d}z\ (\partial\gamma(z))^{-1} (\partial \gamma(w))^{-1} \gamma(z)^{m+1} \gamma(w)^{n+1} \nonumber\\
&\qquad\qquad\qquad\qquad\times\left(\frac{c^\text{m}/2}{(z-w)^4}+\frac{2\, T^\text{m}(w)}{(z-w)^2}+\frac{\partial T^\text{m}(w)}{z-w}\right) \\
&=\oint_0 \mathrm{d}w\ \bigg(\frac{c^\text{m}}{12} \partial^3\big(\gamma^{m+1}(\partial \gamma)^{-1}\big) \gamma^{n+1}(\partial \gamma)^{-1}+2 \partial \big(\gamma^{m+1}(\partial \gamma)^{-1}\big) \gamma^{n+1}(\partial \gamma)^{-1} T^\text{m}\nonumber\\
&\qquad\qquad\qquad\qquad+\gamma^{m+n+2} (\partial \gamma)^{-2} \partial T^\text{m} \bigg) \\
&=\oint_0 \mathrm{d}w\ \bigg(\frac{c^\text{m}}{12} \partial^3\big(\gamma^{m+1}(\partial \gamma)^{-1}\big) \gamma^{n+1}(\partial \gamma)^{-1}+(m-n) \gamma^{m+n+1}(\partial \gamma)^{-1} T^\text{m}\bigg)\\
&=(m-n)\mathcal{L}_{m+n}^\text{m}+ \frac{c^\text{m}}{12}m(m^2-1)\oint_0 \mathrm{d}w\ \gamma^{m+n-1}\partial \gamma\nonumber\\
&\qquad\qquad+\frac{c^\text{m}}{12}\oint_0 \mathrm{d}w\ \partial \left(\gamma^{m+n+2}\left(\frac{3}{2}(\partial^2\gamma)^2 (\partial \gamma)^{-4}-\partial^3 \gamma(\partial \gamma)^{-3}\right)\right)\ .
\end{align}
The last term vanishes upon integration. In the second term, the same integral which defined the identity, see eq.\ \eqref{Idef}, appears. 
Thus we obtain indeed a Virasoro algebra. When identifying $\mathcal{I}=w\mathds{1}$, we see that the central charge equals $c^\text{m}w$, but the generators are again fractionally moded as in the discussion of Section~\ref{subsec:locality and moding}.

\subsection{Higher spin fields}
It should now be clear how to generalise the discussion to an arbitrary chiral field on the world-sheet. Given a primary field $W^{(s)}(z)$ of spin $s$ on the worldsheet, we define its spacetime analogue as 
\be \label{WsDDF}
\mathcal{W}^{(s)}_m\equiv\oint \mathrm{d}z\  \big((\partial \gamma)^{1-s} \gamma^{m+s-1} W^{(s)}\big)(z)\ .
\ee
Note that this expresses basically the coordinate transformation $z \mapsto \gamma(z)$ of the field $W^{(s)}(z)$, except that the coordinate transformation itself is described by a dynamical field. 
The commutation relations of  $\mathcal{W}^{(s)}_m$ can be computed as follows. First, we note that we can organise the OPE of $W^{(s)}(z)$ with another primary $V^{(t)}(w)$ in terms of Virasoro representations, i.e.\ that we can restrict ourselves to the primary fields appearing in the OPE. Thus,
\be 
[\mathcal{W}^{(s)}_m,\mathcal{V}^{(t)}_n]=\sum_{u\ge 0}\oint \mathrm{d}z \Big(c_u(m,n) \, (\partial \gamma)^{1-u} \gamma^{m+u-1} U^{(u)}(z)\Big)+ \text{descendants}\ ,
\ee
where $c_u(m,n)$ are at this stage arbitrary coefficients, and we have summed over all primary fields of spin $u$ appearing in the OPE on the right hand side. The fields $U^{(u)}(z)$ have to appear in this combination with $\gamma$, since this is the only combination which is primary and of conformal weight one on the worldsheet. Moreover, the exponent of $\gamma$ is fixed by noting that $\gamma$ carries charge $-1$ under the $\mathfrak{sl}(2,\mathds{R})$ on the worldsheet and hence the number of $\gamma$'s has to be conserved in the expression. 

In order to determine the $c_u(m,n)$, we note that they do not depend on $\gamma$, and hence we can  compute them by setting $\gamma(z)=z$. Then $\mathcal{W}^{(s)}_m$ agrees with the modes $W^{(s)}_m$ of the worldsheet field, and thus the coefficients are exactly the same as those that appear in the commutation relations of the algebra on the worldsheet. We have therefore shown that 
the commutation relations of the spacetime algebra are identical to the commutation relations of the modes of the respective fields on the worldsheet.

The argument we have presented and the definition \eqref{WsDDF} holds in particular also for the identity field, in which case it reduces to the definition of $\mathcal{I}$, see eq.~\eqref{Idef}. Thus, the central terms of the commutation relations are replaced by $\mathcal{I}$, which can be identified with the spectral flow parameter $w$, as discussed in the main text, see Section~\ref{sec:Identify}.

One can also check that these generators transform indeed as primary fields of spin $s$ in spacetime,
\be 
[\mathcal{L}_m,\mathcal{W}^{(s)}_n]=\big(m(s-1)-n\big) \, \mathcal{W}^{(s)}_{m+n}\ , \label{eq:spacetime primary}
\ee
where $\mathcal{L}_m$ is the spacetime Virasoro algebra \eqref{eq:spacetime Virasoro algebra bos}.

\section{Various commutation relations} \label{app:OPEs}
\subsection{The RNS formalism of strings on \texorpdfstring{$\mathrm{AdS}_3 \times \mathrm{S}^3 \times \mathbb{T}^4$}{AdS3xS3xT4}} \label{subapp:RNS formalism AdS3xS3xT4}
The bosonic part of the worldsheet algebra is $\mathfrak{sl}(2,\mathds{R})_{k+2} \oplus \mathfrak{su}(2)_{k-2}$, together with four free bosons. They have commutation relations
\begin{subequations}
\begin{align}
[\mathscr{J}^3_m,\mathscr{J}^3_n]&=-\tfrac{k+2}{2}m \delta_{m+n,0}\ , \\
[\mathscr{J}^3_m,\mathscr{J}^\pm_n]&=\pm \mathscr{J}^\pm_{m+n}\ , \\
[\mathscr{J}^+_m,\mathscr{J}^-_n]&=(k-2)m \delta_{m+n,0}-2\mathscr{J}^3_{m+n,0}\ , \\
[\mathscr{K}^3_m,\mathscr{K}^3_n]&=\tfrac{k-2}{2}m \delta_{m+n,0}\ , \\
[\mathscr{K}^3_m,\mathscr{K}^\pm_n]&=\pm \mathscr{K}^\pm_{m+n}\ , \\
[\mathscr{K}^+_m,\mathscr{K}^-_n]&=(k+2)m \delta_{m+n,0}+2\mathscr{K}^3_{m+n,0}\ , \\
[\partial X^\alpha_m,\partial \bar{X}^\beta_n]&=m \epsilon^{\alpha\beta} \delta_{m+n,0}\ .
\end{align}
\end{subequations}
There are moreover ten fermions on the worldsheet, which we denote by $\psi^a$, $\chi^a$, $\lambda^\alpha$ and $\bar{\lambda}^\alpha$; they obey the anticommutation relations
\begin{subequations}
\begin{align}
\{\psi^3_r,\psi^3_s\}&=-\tfrac{k}{2}\delta_{r+s,0}\ ,\\
\{\psi^+_r,\psi^-_s\}&=k \delta_{r+s,0}\ ,\\
\{\chi^3_r,\chi^3_s\}&=\tfrac{k}{2}\delta_{r+s,0}\ ,\\
\{\chi^+_r,\chi^-_s\}&=k \delta_{r+s,0}\ ,\\
\{\lambda^\alpha_r,\bar{\lambda}^\beta_s\}&=\epsilon^{\alpha\beta} \delta_{r+s,0}\ .
\end{align}
\end{subequations}
\subsection{The \texorpdfstring{$\mathfrak{psu}(1,1|2)_k$}{psu(1,1|2)k} WZW-model} \label{subapp:psu112k WZW-model}
The $\mathfrak{psu}(1,1|2)_k$ current algebra takes the following form in our conventions:
\begin{subequations}
\begin{align} 
[J^3_m,J^3_n]&=-\tfrac{1}{2}km\delta_{m+n,0}\ ,  \label{eq:psu112 commutation relations a}\\
[J^3_m,J^\pm_n]&=\pm J^\pm_{m+n}\ , \label{eq:psu112 commutation relations b}\\
[J^+_m,J^-_n]&=km\delta_{m+n,0}-2J^3_{m+n}\ , \label{eq:psu112 commutation relations c}\\
[K^3_m,K^3_n]&=\tfrac{1}{2}km\delta_{m+n,0}\ , \label{eq:psu112 commutation relations d}\\
[K^3_m,K^\pm_n]&=\pm K^\pm_{m+n}\ , \label{eq:psu112 commutation relations e}\\
[K^+_m,K^-_n]&=km\delta_{m+n,0}+2K^3_{m+n}\ , \label{eq:psu112 commutation relations f}\\
[J^a_m,S^{\alpha\beta\gamma}_n]&=\tfrac{1}{2}c_a\tensor{(\sigma^a)}{^\alpha_\mu} S^{\mu\beta\gamma}_{m+n}\ , \label{eq:psu112 commutation relations g}\\
[K^a_m,S^{\alpha\beta\gamma}_n]&=\tfrac{1}{2}\tensor{(\sigma^a)}{^\beta_\nu} S^{\alpha\nu\gamma}_{m+n}\, , \label{eq:psu112 commutation relations h}\\
 \{S^{\alpha\beta\gamma}_m,S^{\mu\nu\rho}_n\}&=km \epsilon^{\alpha\mu}\epsilon^{\beta\nu}\epsilon^{\gamma\rho}\delta_{m+n,0}-\epsilon^{\beta\nu}\epsilon^{\gamma\rho} c_a\tensor{(\sigma_a)}{^{\alpha\mu}} J^a_{m+n}\nonumber\\
 &\hspace{6cm}+\epsilon^{\alpha\mu}\epsilon^{\gamma\rho} \tensor{(\sigma_a)}{^{\beta\nu}} K^a_{m+n}\ . \label{eq:psu112 commutation relations i}
\end{align}
\end{subequations}
Here, $\alpha,\beta,\dots$ are spinor indices and take values in $\{+,-\}$. The third spinor index of the supercharges encodes the transformation properties under the outer automorphism $\mathfrak{su}(2)$ of $\mathfrak{psu}(1,1|2)$. Furthermore, $a$ is an $\mathfrak{su}(2)$ adjoint index and takes values in $\{+,-,3\}$. 
The constant $c_a$ equals $-1$ for $a=-$, and $+1$ otherwise. Finally, the $\sigma$-matrices are explicitly given by 
\begin{subequations}
\begin{align}
\tensor{(\sigma^-)}{^+_-}&=2\ , & \tensor{(\sigma^3)}{^-_-}&=-1\ , & \tensor{(\sigma^3)}{^+_+}&=1\ , & \tensor{(\sigma^+)}{^-_+}&=2\ , \label{eq:sigma matrices conventions uud}\\
\tensor{(\sigma_-)}{^{--}}&=1\ , & \tensor{(\sigma_3)}{^{-+}}&=1\ , & \tensor{(\sigma_3)}{^{+-}}&=1\ , & \tensor{(\sigma_+)}{^{++}}&=-1\ , \label{eq:sigma matrices conventions duu} \\
\tensor{(\sigma^-)}{_{--}}&=2\ , & \tensor{(\sigma^3)}{_{+-}}&=1\ , & \tensor{(\sigma^3)}{_{-+}}&=1\ , & \tensor{(\sigma^+)}{_{++}}&=-2\ ,
\label{eq:sigma matrices conventions udd}
\end{align}
\end{subequations}
while all the other components vanish. 
\subsection{The \texorpdfstring{$\mathcal{N}=4$}{N=4} Liouville fields} \label{subapp:N4 Liouville conventions}
To construct $\mathcal{N}=4$ Liouville theory, we start with a free field construction of the $\mathcal{N}=4$ algebra in terms of the (free) fields
\be 
\partial \phi\ , \qquad \psi^{\alpha\beta}\ , \qquad J^a\ ,
\ee
where $\partial\phi$ is a free boson, $\psi^{\alpha\beta}$ are four free fermions, and $J^a$ is a $\mathfrak{su}(2)_{\kappa-1}$ current. They have defining OPEs
\begin{subequations}
\begin{align}
\partial \phi(z)\partial \phi(w)&\sim \frac{1}{(z-w)^2}\ , \\
\psi^{\alpha\beta}(z)\psi^{\gamma\delta}(w)&\sim \frac{\epsilon^{\alpha\gamma}\epsilon^{\beta\delta}}{z-w}\ , \\
J^3(z)J^3(w)&\sim \frac{\kappa-1}{2(z-w)^2}\ , \\
J^3(z)J^\pm(w)&\sim \frac{J^\pm(w)}{z-w}\ , \\
J^+(z)J^-(w)&\sim \frac{\kappa-1}{(z-w)^2}+\frac{2J^3(w)}{z-w}\ .
\end{align}
\end{subequations}
These fields can be organised as an $\mathcal{N}=4$ superfield.
  
 \section{Free field systems}\label{app:freeconv}
 
 \subsection{\texorpdfstring{$bc$}{bc} system}
 
 A $bc$ system consists of  anti-commuting fields $b(z)$ and $c(z)$ with conformal dimensions
 \be
 h(b) = \lambda \ , \qquad h(c) = 1 - \lambda \ , 
 \ee
 and defining OPE
 \be
 b(z) c(w) \sim \frac{1}{z-w} \ . 
 \ee
 The corresponding energy momentum tensor is 
 \be
 T =  (\partial b) c - \lambda\,  \partial ( bc) \ , 
 \ee
 and it has central charge $c=1-3(2\lambda-1)^2$. In particular, the $bc$ system with $\lambda=2$ has $c=-26$. 
 
 \subsection{\texorpdfstring{$\beta\gamma$}{beta gamma} system}
 
 A $\beta\gamma$ system consists of commuting fields $\beta(z)$ and $\gamma(z)$ with conformal dimensions
 \be
 h(\beta) = \lambda \ , \qquad h(\gamma) = 1 - \lambda \ , 
 \ee
 and defining OPE
 \be
 \beta(z) \gamma(w) \sim - \frac{1}{z-w} \ . 
 \ee
 The corresponding energy momentum tensor is 
 \be
 T =   (\partial \beta) \gamma - \lambda\, \partial (\beta \gamma) \ , 
 \ee
 and it has central charge $c=3(2\lambda-1)^2-1$. In particular, the $\beta \gamma$ system with $\lambda=\frac{3}{2}$ has $c=11$. 

 \subsection{Free bosons}
 
 For a free boson with OPE
 \be
 \phi(z) \phi(w) \sim \epsilon \log (z-w) \ , 
 \ee
 and background charge $Q$, the stress-energy tensor is 
 \be
 T = \epsilon \Bigl[ \frac{1}{2} (\partial \phi)^2 - \frac{1}{2} Q \partial^2 \phi \Bigr] \ , 
 \ee
 with central charge
 \be
 c = 1 -3 \epsilon Q^2 \ . 
 \ee
 The conformal dimension of the field $e^{q\phi}$ is then 
 \be
 h\bigl( e^{q\phi} \bigr) = \frac{1}{2} \epsilon q ( q+Q) \ . 
 \ee

\end{document}